\RequirePackage{ifluatex}
\documentclass
[twocolumn,aps,prd,amsmath,amssymb,nofootinbib,floatfix]
{revtex4-1}
\usepackage{CJK}                     
\usepackage[dvips]{graphicx}         
\usepackage{physics}
\usepackage{mathptmx}                
\usepackage{dcolumn}                 
\usepackage{multirow}                
\usepackage{xcolor}                  %
\usepackage{hyperref}
\hypersetup{
 colorlinks=true,     
 linkcolor=blue,      
 citecolor=cyan,      
 filecolor=magenta,   
 urlcolor=cyan         
}
\graphicspath{
{Figures/}
}

\begin{document}

\def\mycom#1{\textcolor{red}{#1}}
\newcommand{\af}[1]{\textcolor{blue}{AF: #1}}
\newcommand{\lr}[1]{\textcolor{red}{LR: #1}}
\newcommand{\fb}[1]{\textcolor{magenta}{FB: #1}}

\def\bea{\begin{eqnarray}} \def\eea{\end{eqnarray}}
\def\be{\begin{equation}} \def\ee{\end{equation}}
\def\bal#1\eal{\begin{align}#1\end{align}}
\def\bse#1\ese{\begin{subequations}#1\end{subequations}}
\def\rra{\right\rangle} \def\lla{\left\langle}
\def\ra{\rightarrow}
\def\rv{\vb{r}}
\def\eps{\epsilon}
\def\ms{\,M_\odot}
\def\mmax{M_\text{max}}
\def\mtov{M_\text{TOV}}
\def\mev{\,\text{MeV}}
\def\fm3{\,\text{fm}^{-3}}
\def\gc3{\,\text{g/cm}^3}
\def\pt{p_\text{th}}
\def\et{\eps_\text{th}}
\def\gt{\Gamma_\text{th}}
\def\FT{Full Table\ }
\def\FT{Finite Temperature\ }
\def\FT{FT\ }
\def\xp{Y_e}  
\def\xp{x_p}  


\title{
Hybrid equation of state approach
in binary neutron-star merger simulations}

\begin{CJK*}{UTF8}{gbsn}

\author{A. Figura$^{1,3}$}
\author{Jia-Jing Lu (陆家靖)$^2$}
\author{G. F. Burgio$^1$}
\author{Zeng-Hua Li (李增花)$^2$}
\author{H.-J. Schulze$^1$}

\affiliation{
$^1$ INFN Sezione di Catania, Dipartimento di Fisica,
Universit\'a di Catania, Via Santa Sofia 64, 95123 Catania, Italy\\
$^2$ Institute of Modern Physics,
Key Laboratory of Nuclear Physics and Ion-beam Application (MOE),
Fudan University, Shanghai 200433, P.R.~China\\
$^3$ Institut f\"ur Theoretische Physik, Universit\"at Frankfurt,
Max~von~Laue~Strasse~1, 60438 Frankfurt, Germany
}

\date{\today}

\begin{abstract}
We investigate the use of hybrid equations of state in binary
neutron-star simulations in full general relativity, where thermal
effects are included in an approximate way through the adiabatic index
$\gt$.
We employ a newly developed finite-temperature equation of state
derived in the Brueckner-Hartree-Fock approach and carry out comparisons
with the corresponding hybrid versions of the same equation of state,
investigating how different choices of $\gt$ affect the
gravitational-wave signal and the hydrodynamical properties of the
remnant. We also perform comparisons with the widely used SFHo equation
of state, detailing the differences between the two cases. Overall, we
determine that when using a hybrid equation of state in binary
neutron-star simulations,
the value of thermal adiabatic index $\gt\approx 1.7$
best approximates the dynamical and thermodynamical
behavior of matter computed using complete, finite-temperature equations
of state.
\end{abstract}


\maketitle
\end{CJK*}

\section{Introduction}

The numerical simulation of neutron star (NS) mergers requires as a most
essential input the equation of state (EOS) of the stellar matter under
the relevant conditions of particle composition, partial densities, and
temperature.

Comparing and contrasting the results of simulations and of the observed
gravitational-wave signal, then allows to constrain theoretical models
for the EOS and extract quantitatively the essential features of the
EOS. The availability of such data has already permitted this selection
process and, in the future, rapid progress is to be expected towards the
identification of ``the'' EOS of dense nuclear matter
\cite{Abbott2017_etal,Abbott2018b}.

Theoretical EOSs have been computed in various approaches, in particular
for cold nuclear matter, but much less for hot matter up to the
temperatures (about 50 MeV) occurring during the merger. In this article
we propose and analyze a finite-temperature EOS derived within the
Brueckner-Hartree-Fock (BHF) many-body approach that has already been
shown to satisfy all current experimental and observational constraints
on nuclear matter \cite{Wei2020}, in particular those imposed by the
merger event GW170817 \cite{Burgio2018,Wei2019}.

We perform here the first binary NS merger simulations with
this EOS and investigate, in particular, the effects of different
approximations for the treatment of finite temperature in the
simulations, following Ref.~\cite{Bauswein:2010dn}. The motivation is to
understand how much the widely used ``hybrid-EOS'' approach impacts the
gravitational-wave properties in binary NS mergers; indeed,
since this approach remains the only viable choice when using a
zero-temperature EOS, it is important to examine which differences are to
be expected with respect to simulations where finite-temperature versions
of the same EOS are employed. In this context, the understanding of the
best setup to be used in the approximate description is of great
importance and can be carried out only by considering the full
temperature-dependent EOS. In particular, we have carried out a number of
simulations of merging NSs in full general relativity,
employing two fully tabulated, temperature dependent EOSs and
neutrino-leakage scheme for the treatment of neutrinos. At the same time,
we have performed similar simulations employing hybrid EOSs whose cold
part is represented by the slice at $T=0$ of the temperature-dependent
EOSs and where we have considered a variety of values for the thermal
adiabatic index $\gt$.
In this way, and summarising the results of a number of simulations,
we conclude that the value of $\gt\approx 1.7$ best approximates the complete,
finite-temperature EOS in binary NS simulations.

The article is organized as follows. We first review in Sec.~\ref{s:eos}
the computation of the EOS in the BHF formalism, with different
approximation for the finite-temperature part. Our numerical setup and
methods are introduced in Sec.~\ref{s:sim}. Results of the simulations
are presented in Sec.~\ref{s:res}, and conclusions are drawn in
Sec.~\ref{s:end}. Technical details regarding the evaluation of
gravitational-wave signal properties are given in the Appendix.

\section{Equation of state at finite temperature}
\label{s:eos}

\subsection{The microscopic BHF approach: the V18 EOS}

We only provide here a brief overview of the formalism, and refer to the
various indicated references for full details, while a more detailed
analysis can be found in \cite{Lu2019}. We here compute the EOS in the
BHF approach for asymmetric nuclear matter at finite temperature
\cite{Baldo1999,Nicotra2006a,Nicotra2006b,Li2010,
  Burgio2011,Burgio2010,Bloch1958,Lejeune1986,Baldo1999a}. The essential
ingredient of this approach is the interaction matrix $K$, which
satisfies the following equations
\begin{widetext}
\begin{eqnarray}
  &&K(n_B,\xp;E) = V + V \, \Re \sum_{1,2}
 \frac{\ket{12} (1-n_1)(1-n_2) \bra{12}}
 {E - e_1-e_2 +i0} K(n_B,\xp;E) \:
\label{eq:BG}
\end{eqnarray}
\end{widetext}
and
\begin{equation}
 U_1(n_B,\xp) = \Re \sum_2 n_2
 \expval{K(n_B,\xp;e_1+e_2)}{12}_a \,,
\label{eq:uk}
\end{equation}
where $n(k)$ is a Fermi distribution, $\xp \equiv n_p/n_B$ is the proton
fraction, and $n_p$ and $n_B$ are the proton and the total baryon number
densities, respectively. (In the following, we will also use the notation
$\rho_i \equiv m_N n_i$ and $\rho \equiv m_N n_B$ for the rest-mass densities,
where $m_N=1.67 \times 10^{-24}\,{\rm g}$ is the nucleon mass). Here, $E$ is
the starting energy and $e(k) \equiv k^2\!/2m + U(k)$ is the single-particle
energy. The multi-indices $1,2$ denote in general momentum, isospin, and
spin. In the present calculations, we adopt the Argonne $V_{18}$
\cite{Wiringa95} potential as a realistic nucleon-nucleon interaction $V$
supplemented with compatible microscopic three-nucleon forces (TBF),
derived by employing the same meson-exchange parameters as the two-body
potential \cite{Zuo2002,Li2008a,Li2008b,Grange1989}.

Regarding the extension to finite temperature, we use the so-called
frozen-correlations approximation \cite{Nicotra2006a, Nicotra2006b,
  Li2010, Burgio2011, Baldo1999a}, and approximate the single-particle
potentials $U_{n,p}(k)$ by the ones calculated at $T=0$. Within this
approximation, the nucleonic free energy density has the following
simplified expression,
\begin{equation}
 f_N = \sum_{i=n,p} \qty[ 2\sum_k n_i(k)
 \qty( {k^2\over 2m_i} + {1\over 2}U_i(k) ) - Ts_i ] \,,
\label{eq:f}
\end{equation}
where
\begin{equation}
 s_i = - 2\sum_k \qty( n_i(k) \ln n_i(k) + \qty[1-n_i(k)] \ln \qty[1-n_i(k)] )\,,
\label{eq:entr}
\end{equation}
is the entropy density for the component $i$ treated as a free Fermi gas
with spectrum $e_i(k)$. From the total free energy density $f=f_N+f_L$
including lepton contributions, all relevant observables can be computed
in a thermodynamically consistent way, namely one defines the chemical
potentials
\begin{equation}
 \mu_i = \pdv{f}{n_i} \,,
\end{equation}
which allow to calculate the composition of betastable stellar matter,
and then the total pressure $p$ and the specific internal energy $\eps$
\begin{eqnarray}
 p &=& n_B^2 \pdv{(f/n_B)}{n_B}
 = \sum_i \mu_i n_i - f \,,
\label{e:eosp}
\\
 \eps &=& \frac{f + Ts}{\rho} \,,\quad
 s = -\pdv{f}{T} \,,
\label{e:eose}
\end{eqnarray}
so that $e\equiv\rho(1+\eps)$ is the total energy density.

In practice, numerical parametrizations for the free energy density of
symmetric nuclear matter (SNM) and pure neutron matter (PNM) were given
in Ref.~\cite{Lu2019}, and for asymmetric nuclear matter a parabolic
approximation for the $\xp$ dependence is used
\cite{Burgio2010,Zuo2004,Bombaci1991,Zuo1999},
\begin{eqnarray}
 f(n_B,T,\xp) &\approx& f_\text{SNM}(n_B,T)
\\&&\nonumber
 + (1-2\xp)^2 \qty[ f_\text{PNM}(n_B,T) - f_\text{SNM}(n_B,T) ]
\,.
\label{e:parab}
\end{eqnarray}
This specifies the EOS for arbitrary values of baryon density, proton
fraction, and temperature, which can then be employed in merger simulations,
or simply for computing the mass-radius relation of cold and hot NSs
by solving the Tolmann-Oppenheimer-Volkov (TOV) equations for
charge-neutral betastable matter including leptons. We also report that the V18 EOS becomes acausal at $n_B=0.75\fm3$ ($\rho \approx 1.3\times 10^{15}$ g/cm$^3$; see, e.g., Ref. \cite{Taranto2013}); this density, however, is far from ever being reached in the simulations (see Fig.~\ref{f:maxrhot} and related discussion).

Since our EOS, which hereafter we refer to as the V18 EOS, accounts only
for homogeneous matter in the core region of the NS, we
properly extend the EOS, for every temperature and proton fraction, with
an EOS for the crust, which we define as that covering the range in
rest-mass densities $\rho \lesssim 10^{14}\gc3$. In particular, we
choose the Shen EOS \cite{Shen11} for that purpose. Furthermore, an
artificial low-density background atmosphere, $\rho \lesssim 10^3\gc3$,
evolved as discussed in \cite{Radice2013c}, is used in all our
simulations.

\subsection{The phenomenological SFHo EOS}

As an alternative to the temperature-dependent V18 EOS and to extend and
strengthen the results of our comparison we have also considered an
alternative temperature-dependent EOS, namely, the phenomenological SFHo EOS
\cite{Hempel2009,Hempel2012}.
We recall that phenomenological approaches are commonly used in
simulations of core-collapse supernovae and NS mergers, where a
wide range of densities, temperatures, and charge fractions, describing
both clustered and homogeneous matter, has to be covered. Some of the
most commonly used finite-temperature EOSs are the ones by Lattimer \&
Swesty \cite{Lattimer91} and Shen et al.~\cite{shen98}. In both cases,
matter is modelled as a mixture of heavy nuclei treated in the
single-nucleus approximation, $\alpha$ particles, and free neutrons and
protons immersed in a uniform gas of leptons and photons. In the former
case, nuclei are described within the liquid-drop model, and a simplified
Skyrme interaction is used for nucleons; in the latter case a
relativistic mean field (RMF) model based on the TM1 interaction \cite{Sugahara1993}
is used for nucleons.
In both approaches, all light nuclei are ignored, except for alpha particles.
This drawback has been overcome in the SFHo EOS model of
Hempel \& Schaffner-Bielich (HS) \cite{Hempel2009}
and Hempel et al.~\cite{Hempel2012},
which goes beyond the single-nucleus approximation,
and takes into account a statistical ensemble of nuclei
and interacting nucleons.
Nuclei are described as classical Maxwell-Boltzmann particles,
and nucleons are treated within the RMF
model employing different parameterizations.

\begin{figure*}[t]
\vspace{-3mm}
\centerline{\includegraphics[width=0.9\textwidth]{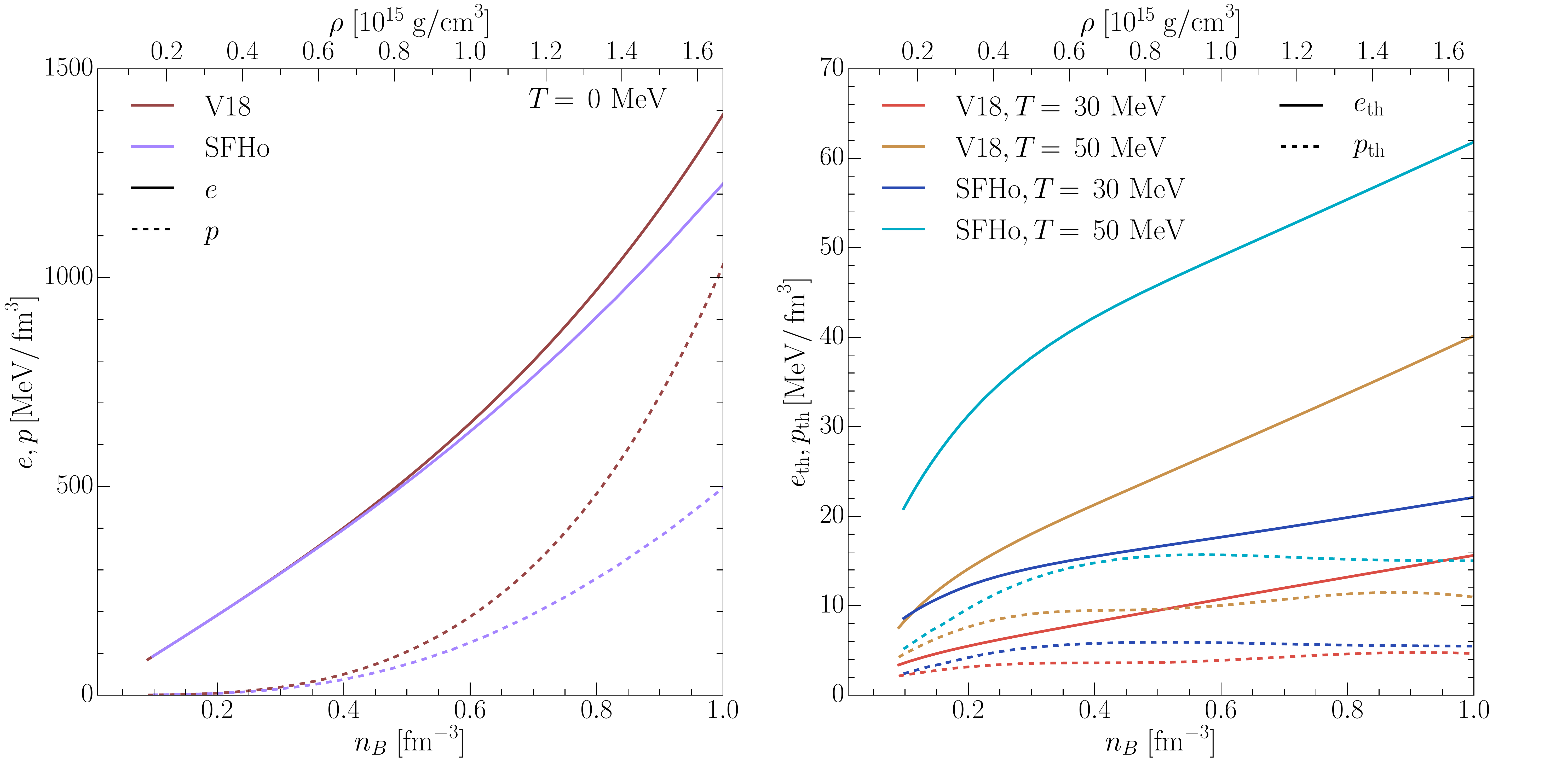}}
\vspace{-3mm}
\caption{Left panel: pressure $p$ and energy density $e$ of betastable
  matter at $T=0$ as a function of the baryon number density. Right
  panel: thermal pressure and internal energy density,
  Eqs.~(\ref{e:pth}), (\ref{e:eth}) at different temperatures. Results
  with V18 and SFHo EOSs are compared.}
\label{f:xt}
\end{figure*}

Here, we adopt the new SFHo EOS \cite{Steiner2013}, which is based on the
HS EOS but implemented with a new RMF parameterization fitted to some
NS radius determinations. The new RMF parameters are varied to
ensure that saturation properties of nuclear matter are correctly
reproduced. In particular, the nuclear incompressibility $K=245\mev$
turns out to be compatible with the currently acceptable range of
$240\pm 20\mev$ \cite{Colo2004}, which agrees with that predicted
from the giant monopole resonances. Moreover, the new parameterization
ensures that the symmetry energy at saturation density $J=32.8\mev$
is well within the empirical range $28.5 - 34.9\mev$
\cite{Burgio2018c}, and that the NS maximum mass
$\mtov=2.06\ms$ is (marginally) compatible with the currently strongest
observational constraint $M>2.14^{+0.10}_{-0.09}\ms$
\cite{Cromartie2019}.

\begin{figure*}[t]
\vspace{-1mm}\hspace{0mm}
\centerline{\includegraphics[width=0.9\textwidth]{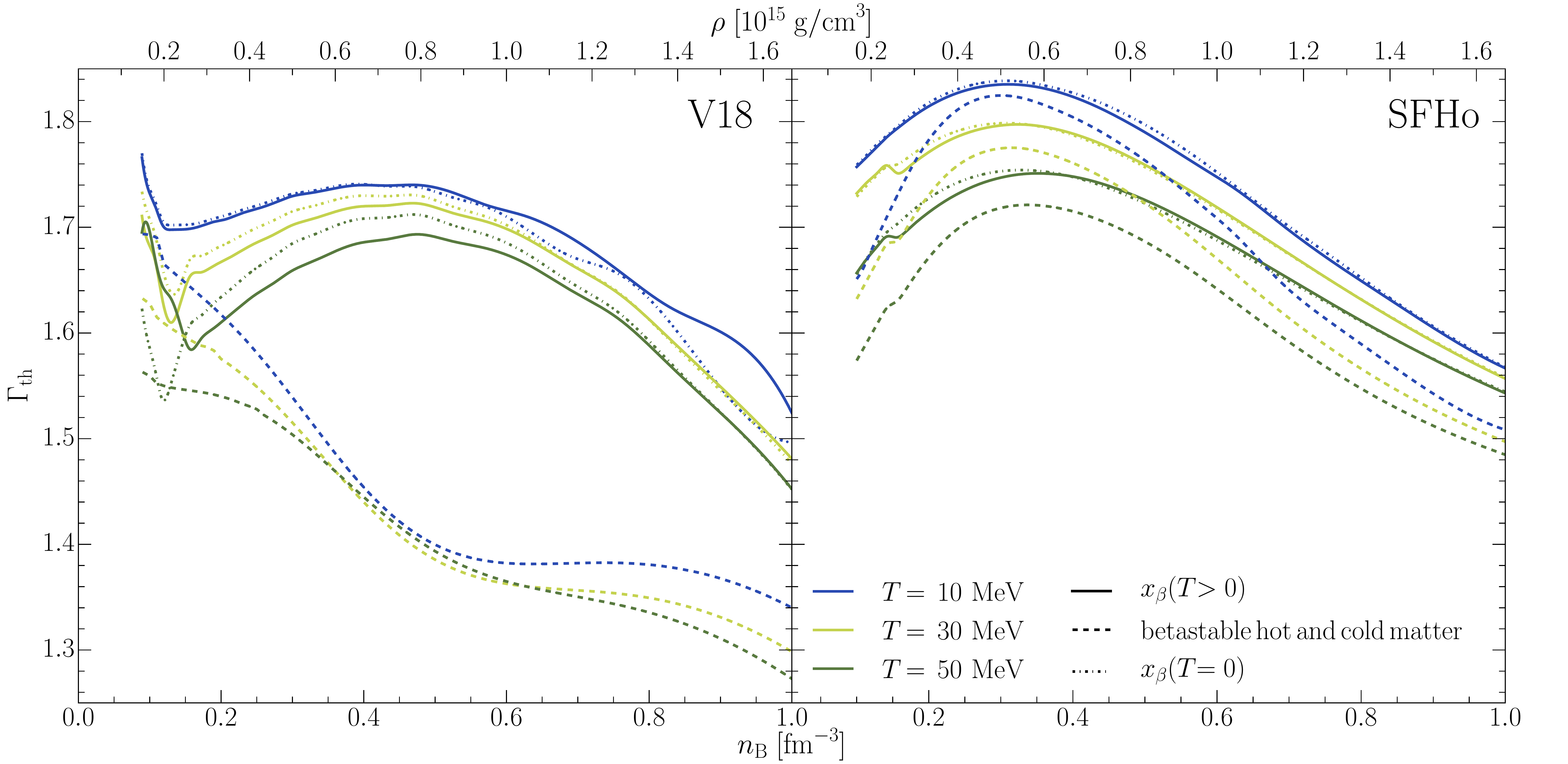}}
\vspace{-6mm}
\caption{Adiabatic index of betastable matter, Eq.~(\ref{e:gth}), as a
  function of density at different temperatures. Dashed curves show
  results obtained from betastable hot and cold matter, while for the
  solid (dash-dotted) curves the proton fraction is fixed to the one of
  betastable hot (cold) matter (see discussion in the text).}
\label{f:gam}
\end{figure*}

As an illustration of the properties of these two temperature-dependent
EOSs, Fig.~\ref{f:xt} shows the pressure $p$ and energy density $e$ of
betastable matter as a function of the baryon number density for both the
V18 and SFHo EOSs. In particular, in the left panel we display the energy
density (solid lines) and pressure (dashed curves) as a function of the
baryon density obtained at $T=0$ for the V18 case, and the SFHo EOS. We
notice that the V18 EOS is stiffer than SFHo and this will play an
important role in the discussion and interpretation of the simulation
results. In the right panel, on the other hand, we display the thermal
contributions to the betastable EOS defined as
\bal
  \pt(\rho,T) &\equiv p(\rho,T)-p(\rho,0) \:,
\label{e:pth}
\\
 e_{\rm th}(\rho,T) &\equiv \rho \qty[\eps(\rho,T)-\eps(\rho,0)] \:,
\label{e:eth}
\eal
for different temperatures ($T=30,50\mev$) and where $e_{\rm th}$
is the internal energy density.
One can notice that in the V18
case the overall thermal effects are smaller than in SFHo, of the order
of a few percent at high density, even at the fairly high temperature
$T=50\mev$ considered here (see \cite{Carbone2019b} for a study on
uncertainties of finite-temperature properties of neutron matter). In
Ref.~\cite{Lu2019} we examined in detail for the V18 case the intricate
interplay between the nucleonic and leptonic contributions to the
betastable EOS, which are of equal importance.

\subsection{Hybrid-EOS approach}
\label{s:hybrid_EOS}

An approach often employed in simulations of NS mergers
\cite{Janka93,Bauswein:2010dn,Baiotti08,Hotokezaka2011,Kiuchi2014,
  DePietri2016,Endrizzi2016,Hanauske2016,Ciolfi2017,Shibata:2017b,
  Radice2018,Radice2018a,Alford2018,Endrizzi2018,Kiuchi2019,DePietri2020}
is the so-called ``hybrid-EOS'', in which pressure and the specific
internal energy can be expressed as the sum of a ``cold'' contribution,
obeying a zero-temperature EOS, and of a ``thermal''
contribution obeying the ideal-fluid EOS
(see \cite{Rezzolla_book:2013} for details).
In this approach, the relation
between the thermal pressure and the internal energy density of
betastable matter can be expressed as
\begin{equation}
\pt(\rho,T) = e_{\rm th} (\gt-1)  \,.
\label{e:gthb}
\end{equation}
where $\gt$ is the thermal adiabatic index appearing in the ideal-fluid
approximation. In a temperature-dependent approach, this quantity becomes
dependent on density and temperature, i.e., $\gt\equiv1+\pt/e_{\rm{th}}$, and
this dependence is illustrated in Fig.~\ref{f:gam} with dashed curves for
the V18 (left panel) and for the SFHo EOS (right panel). Note that there
is a clear density dependence, whereas the temperature dependence turns
out to be less pronounced. Overall, the thermal adiabatic index remains
above $1.5$ at all densities in the SFHo case, but decreases below 1.5 in
the V18 case, consistent with the thermal pressures shown in
Fig.~\ref{f:xt}.

In temperature-dependent EOSs to be used in numerical simulations, the
adiabatic index is usually not defined for betastable matter (featuring
different proton fractions in hot and cold matter), but can be computed
at constant proton fraction as
\begin{equation}
 \gt(\rho,T) \equiv
 1 + \frac{p(\rho,x_\beta,T)-p(\rho,x_\beta,0)}
          {\rho\qty[\eps(\rho,x_\beta,T)-\eps(\rho,x_\beta,0)]} \,,
\label{e:gth}
\end{equation}
where $x_\beta$ is the betastable proton fraction at either ($\rho,T>0$)
or ($\rho,T=0$). This leads to different numerical values that are also
displayed in Fig.~\ref{f:gam}, where the solid (dash-dotted) curves
display results with $x_\beta$ taken at $T>0\,(T=0)$ for the V18 (left
panel) and the SFHo EOS (right panel), respectively. We note that this
procedure yields values $1.5 \lesssim \gt \lesssim 1.7$ for the V18 EOS,
and $1.6 \lesssim \gt \lesssim 1.8$ for the SFHo EOS, whereas the average
value for the betastable matter is smaller in both cases. We point out,
however, that in the merger simulations the matter in the early remnant
is usually not in beta equilibrium and therefore all the values shown in
Fig.~\ref{f:gam} can only give qualitative indications of effective $\gt$
values. This will be discussed in more detail later.

In fact, three-dimensional hydrodynamical calculations of NS
mergers in the conformally flat approximation of general relativity
reported in Ref.~\cite{Bauswein:2010dn} have questioned the validity of a
constant-$\gt$ approximation in the hybrid-EOS approach (originally
chosen as $\gt\approx1.5$ \cite{Janka93}), especially in the postmerger
phase, where thermal effects are most relevant. Strong variations were
found in both the oscillation frequency of the forming hypermassive
NS (HMNS), and the delay time between the merger and black-hole
formation, with respect to the simulations with a fully consistent
treatment of the temperature. It is one of our goals here to reconsider
-- by comparing and contrasting fully general-relativistic simulations
with temperature-dependent and hybrid EOSs -- the issue of the most
appropriate constant value of $\gt$ to be employed when
adding a thermal component to the EOS.

\begin{figure*}[t]
\vspace{-3mm}
\centerline{\hspace{0mm}\includegraphics[width=0.9\textwidth]{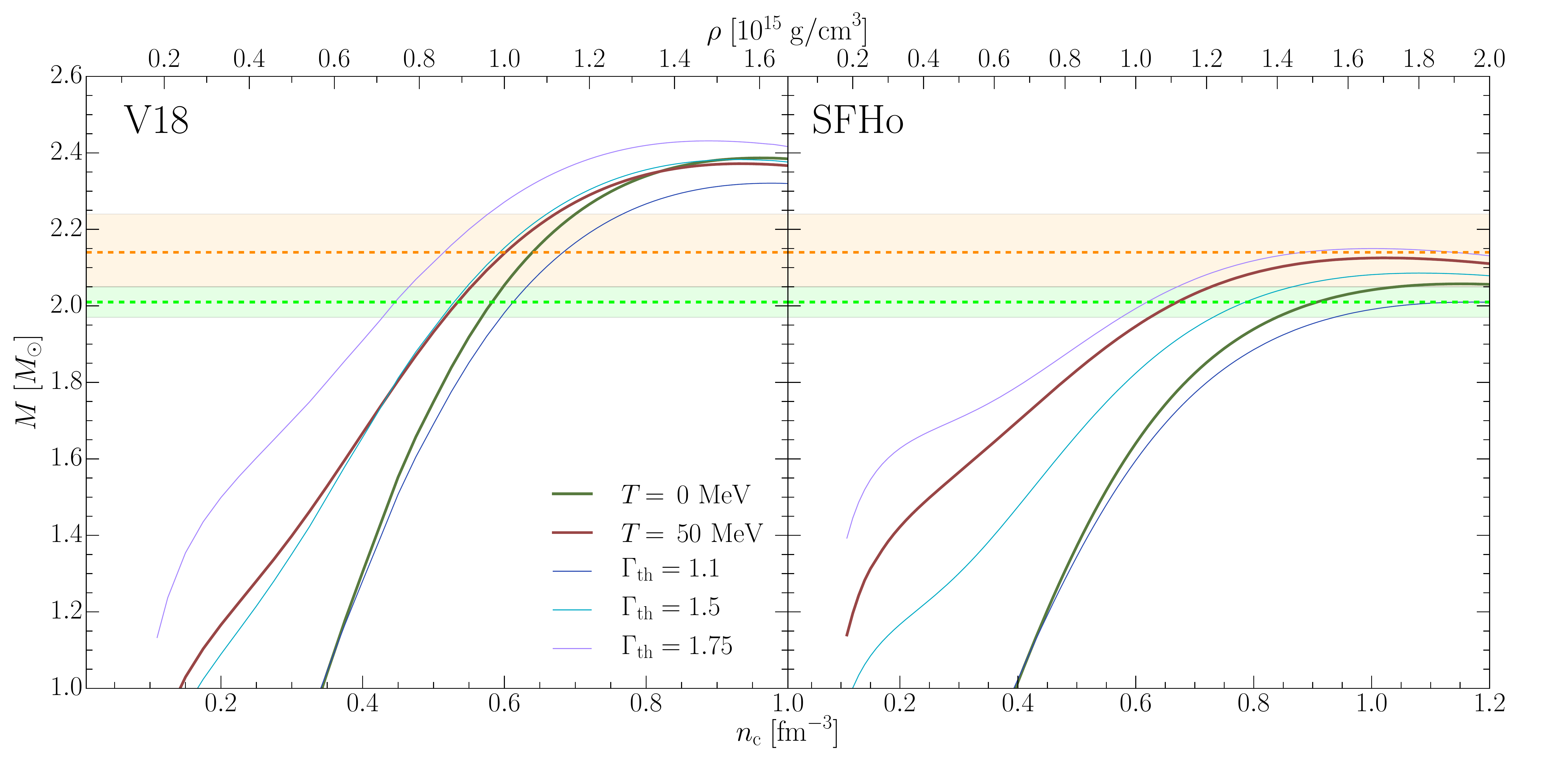}}
\vspace{-3mm}
\caption{
Gravitational mass as a function of the central rest-mass density relations
for $T=0$ and $T=50\mev$ with full temperature treatment,
and different choices of constant $\gt=1.1, 1.5, 1.75$ at $T=50\mev$.
Dashed orange and green lines,
together with the shaded regions of the same color,
refer to the observational constraints of
Refs.~\cite{Cromartie2019,Antoniadis2013}, respectively.}
\label{f:mrho}
\end{figure*}

\subsection{Macroscopical properties of the V18 and SFHo EOSs}
\label{s:gt}

Given the widespread recent use of hybrid EOSs
\cite{Kiuchi2014,Takami2014,Takami2015,Rezzolla2016,
  DePietri2016,Endrizzi2016,Hanauske2016,Ciolfi2017,Shibata:2017b,
  Radice2018,Radice2018a,Alford2018,Endrizzi2018,Kiuchi2019,DePietri2020}
and the scarcity of fully temperature dependent EOSs (that are
effectively restricted to a handful
\cite{Banik2014,Typel2010,Steiner2013,Hempel2012,Togashi2016,Most2019c}),
the determination of the most realistic value to be used for
$\gt$ is not purely academic. Indeed, even at the
lowest-order approximation, $\gt$ has an impact on the
stability of the merger remnant and hence of its lifetime before
collapsing to a black hole. This is most easily shown in
Fig.~\ref{f:mrho}, which reports sequences of nonrotating equilibrium
models as a function of the central rest-mass density (or baryon number
density) for the V18 EOS (left panel) and the SFHo EOS (right
panel). Different curves refer to different temperatures (i.e., $T=0$ and
$T=50\mev$), using the exact temperature dependence and  three
different choices of constant $\gt=1.1, 1.5, 1.75$. In other words, we
use Eq.~\eqref{e:eth} at $T=50\mev$
and the estimate of the thermal adiabatic index
to compute the thermal contribution to the pressure,
Eq.~\eqref{e:gthb}
\footnote{
For this plot, a cold crust is attached
to the isothermal NS interior
at $n_B=0.08\fm3$,
corresponding to $\rho \approx 1.32 \times 10^{14}\gc3$.
}.

Note the weak dependence of the maximum TOV mass on the temperature, so that
for the V18 EOS we have that $\mtov(T=0) \equiv \mtov=2.387\ms$ at a central
rest-mass density $\rho_c=1.58\times10^{15}\gc3$
(corresponding to a baryon number density $n_c=0.96\fm3$), while
$\mtov(T=50\mev)=2.372\ms$ at
$\rho_c=1.53\times10^{15}\gc3$ ($n_c=0.93\fm3$).
This is mainly due to the competition of three
different effects for fixed density and increasing temperature, namely
a) the increase of the thermal pressures
of neutrons and protons,
b) the increase of the isospin symmetry due to beta-stability,
which reduces the baryonic pressure, and
c) the increase of the lepton thermal pressure.  In particular,
the V18 EOS is characterized by large values of the symmetry energy
which increases with temperature and density,
and this is due to the strongly repulsive character
of the microscopic three-body forces.
This implies a strong increase of the isospin symmetry
with temperature and density \cite{Lu2019}.

In Fig.~\ref{f:mrho}, left panel,
the $\gt=1.5$ approximation at $T=50\mev$ happens to yield a very
similar result as the full calculation,
hence we can conclude that for the V18 EOS,
the value of the adiabatic thermal index $\gt = 1.5$ represents the
best approximation for betastable matter at finite temperature
as it is the one that best mimics the effects of a full temperature
dependence. The proton fraction $x_p$ corresponding to betastability is quite different at $T=0$ and finite $T$ at given baryon density, and therefore the $\gt$ computed in this way is different from the one calculated at the same $x_p$ in both cold and hot matter, using either the $x_p$ of cold matter or the one of hot matter in Eq. (\ref{e:gth}). The latter is the choice made in the numerical simulations and, according to Fig. \ref{f:gam}, typical values of $\gt \sim 1.7$ in this choice correspond to typical values of $\gt \sim 1.5$ in the betastable procedure, which is the one used in Fig.~\ref{f:mrho}.

On the other hand, $\gt=1.1$ and 1.75 predict lower and higher $\mtov$,
respectively, according to the lower and higher thermal pressure they provide.
One can appreciate the opposite effects of $\pt$ and $e_{\rm th}$
on the maximum mass:
when including only $e_{\rm th}$
($\gt=1.1$ curves featuring very small $p_{\rm th}$),
$\mtov$ decreases with respect to the cold $\mtov (T=0)$,
whereas including also $\pt$ (FT, $\gt$=1.5, 1.75 curves)
$\mtov$ increases again.
For the V18 FT EOS there is nearly compensation between both effects
due to a relatively low thermal pressure, induced
by a strong change of the proton fraction in hot vs cold matter,
and the related changes of hadronic and leptonic contributions
to the pressure that compete with each other, as explained before.

The right panel of Fig.~\ref{f:mrho} reports the corresponding results
for the SFHo case,
and in this case we can note a larger temperature dependence of the
maximum TOV mass when compared to the V18 case;
in turn, this relates to the higher thermal pressure and adiabatic index
for the SFHo.
This is due to the smaller change of the proton fraction
with increasing temperature,
which causes a larger thermal pressure,
at variance with the V18 EOS case.
Consequently,
the full calculation at $T=50\mev$ seems to be better reproduced
by the $\gt\simeq 1.7$ approximation here.

The maximum masses are then $\mtov(T=0) \equiv \mtov=2.058\ms$,
with a central rest mass density
$\rho_c = 1.90\times10^{15}\gc3$ ($n_c=1.15\fm3$), and
$\mtov(T=50\mev)=2.126\ms$, with
$\rho_c = 1.68\times10^{15}\gc3$ ($n_c=1.02\fm3$).
These values, together with other useful information such as the
rotation frequencies at the mass-shedding limit,
are summarized in Table~\ref{t:eos}.

\begin{table}[t]
\caption{
Properties of the maximum mass configurations of both static
and maximally rotating stars with Kepler frequency at temperatures
$T=0$ and $50\mev$:
gravitational and baryonic masses $M$ and $M_B$,
and the equatorial radius $R$.}
\def\myc#1{\multicolumn{1}{c|}{$#1$}}
\setlength{\tabcolsep}{2.4pt}
\renewcommand{\arraystretch}{1.2}
\begin{ruledtabular}
\begin{tabular}{lccccc}
 EOS & $f$  & $T$   & $M$     & $M_B$   & $R$  \\
     & [Hz] & [MeV] & [$\ms$] & [$\ms$] & [km] \\
\hline\multirow{4}{*}{V18}
  &  0     &  0    &  2.387    &  2.913    &  10.86  \\
  &  0     &  50   &  2.372    &  2.785    &  11.40  \\
  &  1770  &  0    &  2.845    &  3.385    &  14.17  \\
  &  1590  &  50   &  2.724    &  3.102    &  15.00  \\
\hline\multirow{4}{*}{SFHo}
  &  0     &  0    &  2.058    &  2.448    &  10.30  \\
  &  0     &  50   &  2.126    &  2.450    &  11.81  \\
  &  1741  &  0    &  2.472    &  2.911    &  13.73  \\
  &  1376  &  50   &  2.413    &  2.726    &  15.98  \\
\end{tabular}
\end{ruledtabular}
\label{t:eos}
\end{table}

Finally, we note that the merger remnant is expected to be rotating
differentially and to support a mass which is upper bounded only by the
threshold mass to prompt collapse to a black hole, that can be estimated
to be \cite{Koeppel2019}
\begin{equation}
 M_\text{th} = \mtov \qty( 3.06 - \frac{1.01}{1-1.34\,\mtov/R_\text{TOV}} )
 \:.
\label{e:mth}
\end{equation}
For the V18 EOS, the threshold mass Eq.~(\ref{e:mth}) amounts to
$M_\text{th} = 3.04\ms$ with $\mtov/R_\text{TOV}=0.324$,
whereas in the SFHo case
$M_\text{th} = 2.86\ms$, being $\mtov/R_\text{TOV}=0.295$
(in geometrized units with $c=1=G=M_\odot$).

\section{Initial data and numerical procedure of merger simulations}
\label{s:sim}

The mathematical and numerical setup considered here is similar to the
one discussed in great detail in Ref.~\cite{Papenfort2018}; we review
here only the main aspects and differences with respect to this
reference, referring the interested reader to the latter for additional
information. We consider initial data for irrotational binary neutron
stars computed using the multi-domain spectral-method code
\texttt{LORENE} \cite{Lorene,Gourgoulhon01}. All initial data have been
modeled considering a zero-temperature, beta-equilibrated cut of the full
EOS table (which will be labeled from now on as ``cold EOS''), and
involve, in our case, equal-masses binaries with a gravitational mass
$M=1.35\ms$ at infinite separation (corresponding to a total baryonic
mass $M_{\rm b}=2.97\ms$ with the V18 EOS and $M_{\rm b}=2.96\ms$ with
the SFHo EOS), and an initial separation between the stellar centers of
45 km.

We then proceed to study two different implementations of our
finite-temperature EOS:

\textit{(a)} The fully-tabulated (FT) case, in which a local temperature
is obtained by inverting the $e(\rho,\xp,T)$ entries in the EOS table,
using the values of the internal energy density $e$, rest-mass density
$\rho$, and proton fraction $\xp$ obtained through the solution of the
hydrodynamics equations at a given timestep. This temperature is then
used to obtain the total pressure $p(\rho,\xp,T)$ from the same EOS
table.

\textit{(b)} The ``hybrid-EOS'' method discussed in
Sec. \ref{s:hybrid_EOS}, where finite-temperature effects, caused in
particular by shock heating during the postmerger phase, are taken into
account by enhancing the zero-temperature EOS with an ideal fluid
contribution \cite{Janka93,Rezzolla_book:2013}. In this method, the local
pressure is approximated by
\begin{equation}
  p = p_c + \rho (\eps-\eps_c)(\gt-1) \,,
\end{equation}
using the values $\eps_c(\rho)$ and $p_c(\rho)$ of the cold EOS table for
{\em betastable} matter and the local propagated values of $\rho$ and
$\eps$. In this case no local temperature (and no proton fraction) can be
extracted during the simulation. The adiabatic index $\gt$ is a constant
both in space and time, constrained mathematically and from first
principles to be $1 \leq \gt \leq 2$ \cite{Carbone2019}. However, in
order to properly compare a simulation of type \textit{(b)} with the
corresponding simulation of type \textit{(a)}, the cold part of the
hybrid EOS is chosen to match the $T=0$ slice of the
temperature-dependent EOS. In this way, the solutions of type
\textit{(a)} and \textit{(b)} are identical during the inspiral -- when
shocks are absent or minute and confined to the stellar surfaces -- but
start to differ after the merger, when thermal effects develop. Clearly,
we consider the simulations of type \textit{(a)} as the most realistic
ones and iterate the values of $\gt$ in simulations of type \textit{(b)}
to find the closest match in the bulk behavior of the matter.

Overall, for our V18 EOS we consider five different binary merger
simulations, namely the reference \FT case
[i.e., one simulation of type \textit{(a)}]
and four additional simulations in which the value of $\gt$ is varied
[i.e., four simulations of type \textit{(b)}].
In particular, we consider the limiting case of $\gt=1.1$,
representative of the ``cold'' case with almost absent thermal effects
-- the case $\gt=1.5$, which best approximates the V18 EOS
in the betastable regime at $T=50\mev$
according to Figs.~\ref{f:gam} and \ref{f:mrho}
-- the case $\gt=1.7$,
which best approximates the \FT results in the simulations
--~and, finally, the case $\gt=1.75$ with the largest thermal contributions,
which also represents a common choice in the literature
(see Refs.~\cite{Bauswein:2010dn,Takami2015} for discussions
on the use of different $\gt$).
In addition, we also perform three more simulations with the SFHo EOS,
one in the \FT regime and two using
the hybrid-EOS approach with $\gt=1.5$ and $\gt=1.75$.

All simulations are performed in full GR using the fourth-order
finite-differencing code of McLachlan \cite{Brown:2008sb}, which is part
of the publicly available Einstein Toolkit \cite{loeffler_2011_et}. The
code solves the CCZ4 formulation of the Einstein equations
\cite{Alic:2011a,Alic2013,Bezares2017}, with a ``1+log'' slicing
condition and a ``Gamma driver'' shift condition (see, e.g.,
Refs.~\cite{Alcubierre2003, Pollney:2007ss:shortal}). The
general-relativistic hydrodynamics equations are solved using the
\texttt{WhiskyTHC} code \cite{Radice2013b,Radice2013c,Radice2015}, which
uses either finite-volume or high-order finite-differencing
high-resolution shock-capturing methods. We employ, in particular, the
local Lax-Friedrichs Riemann solver (LLF) and the high-order MP5
primitive reconstruction \cite{suresh_1997_amp,Radice2012a}. For the time
integration of the coupled set of hydrodynamic and Einstein equations we
use the method of lines with an explicit third-order Runge-Kutta method,
with a Courant-Friedrichs-Lewy (CFL) number of $0.15$ to compute the timestep.

Although matter compression and shocks increase the temperature of the
remnant to several tens of MeV \cite{Radice:10}, neutrino emission acts
as cooling mechanism and is implemented in the temperature-dependent
simulations as only in the latter the electron fraction is consistently
evolved in time. In these cases, we treat the effects on matter due to
weak reactions using the gray (energy-averaged) neutrino-leakage scheme
described in Refs. \cite{Galeazzi2013,Radice2016}, and evolve
free-streaming neutrinos according to the ${\rm M0}$ heating scheme
introduced in Refs. \cite{Radice2016,Radice2018a}.

To ensure the non-linear stability of the spacetime evolution, we add a
fifth-order Kreiss-Oliger-type artificial dissipation \cite{Kreiss73}. We
employ an adaptive-mesh-refinement approach, where the grid hierarchy is
handled by the Carpet driver \cite{Schnetter-etal-03b}. Such a hierarchy
consists of six refinement levels with a grid resolution varying from
$h_5 = 0.16\ms$ (i.e., $\sim 236\,{\rm m}$) for the finest level,
corresponding to about 40 points covering the NS radius on the equatorial plane
at the beginning of the simulation for both the V18 and SFHo models,
to $h_0 = 5.12\ms$ (i.e., $\sim 7.5\,{\rm km}$) for the coarsest level,
whose outer boundary is at $1024\ms$ (i.e., $\sim 1515\,{\rm km}$).
To reduce computational costs,
we adopt a reflection symmetry across the $z=0$ plane.
While the V18 simulations presented here follow the remnant
evolution for a timescale of at least $20\,{\rm ms}$, the SFHo
simulations are stopped a few milliseconds after the collapse to a black hole.

Before concluding this Section, a couple of remarks are useful. First,
the hybrid-EOS simulations are carried out using the betastable tables at
$T=0$, so that the simulation is ``forced" to treat betastable matter,
corrected with the already-described finite-temperature effects. The \FT
simulation, on the other hand, is free to drive away from the betastable
condition, and indeed this is what happens starting from the very
beginning, as we will discuss in the next Section.
Second, the simulations employing the V18 EOS discussed here
represent the first application of such recently derived and publicly
available temperature-dependent EOS \cite{V18EOS}.

\begin{figure*}[t]
\vspace{-0mm}
\centerline{\hspace{5mm}\includegraphics[scale=0.31]{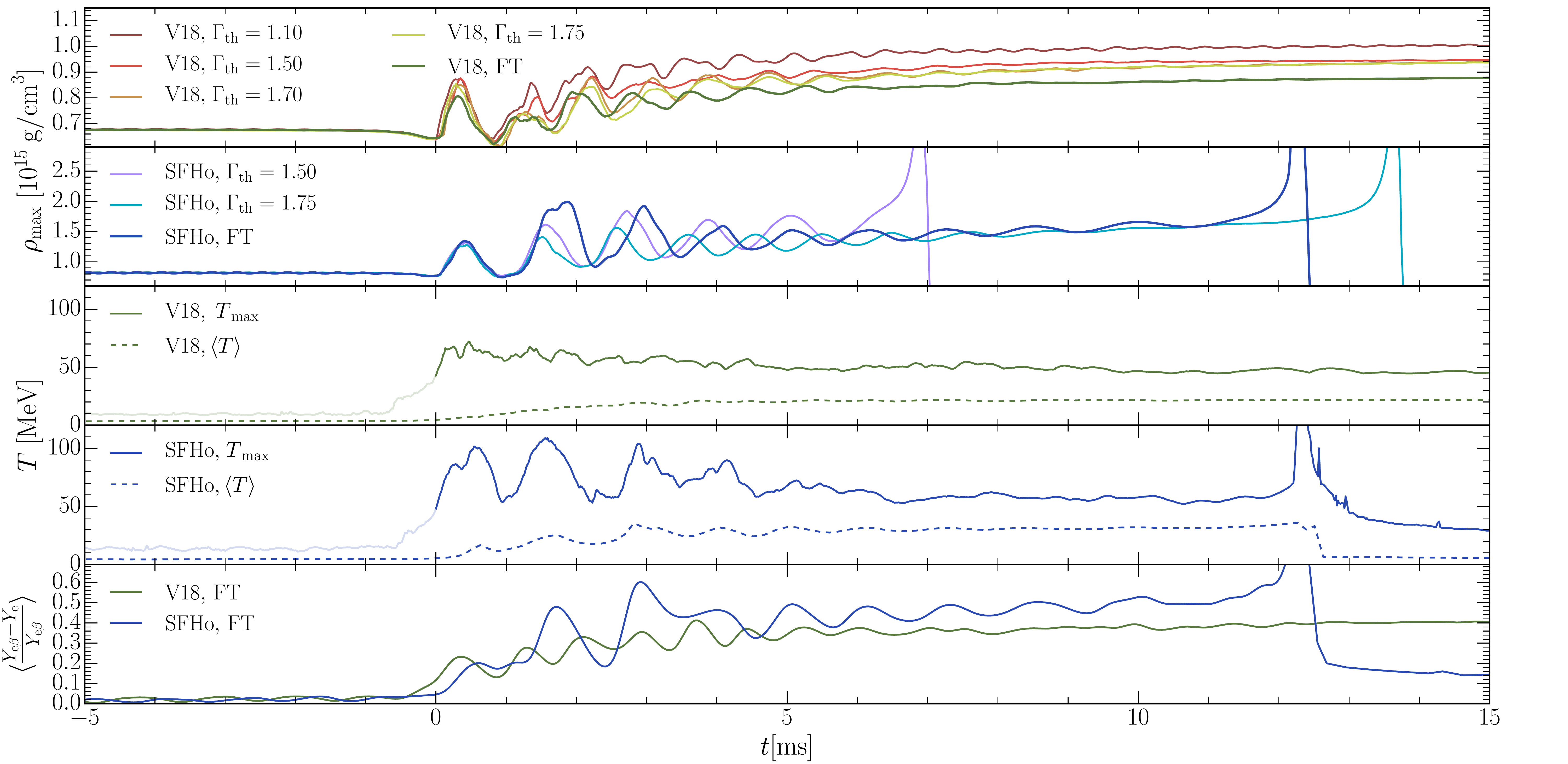}}
\vspace{-3mm}
\caption{
Maximum values of rest-mass density (upper panels) and temperature
(third and fourth panels from the top,
only for the simulations using the FT EOSs)
as a function of time.
The evolution of the average temperature $\expval{T}$, Eq.~(\ref{e:tav}),
is also displayed.
A lighter color is chosen for the inspiral phase,
where such temperatures are meant as representative only and do not reflect
an accurate description of the thermodynamics of the matter.
The average deviation from beta-stability, Eq.~(\ref{e:dyav}),
is also represented in the lowest panel for both FT EOSs.}
\label{f:maxrhot}
\end{figure*}

\begin{figure*}[t]
\vspace{-0mm}
\centerline{\hspace{7mm}\includegraphics[scale=0.31]{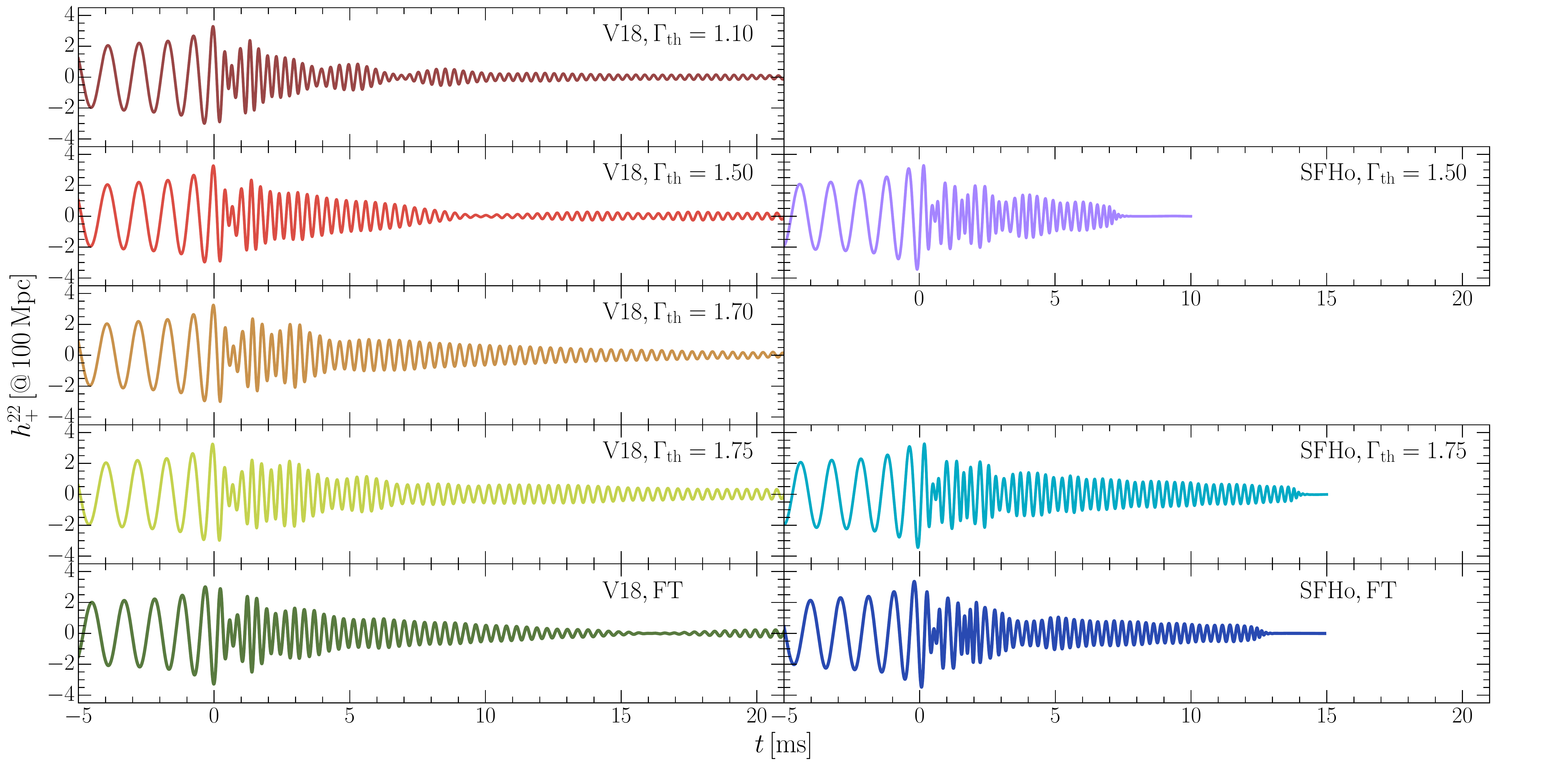}}
\vspace{-3mm}
\caption{ Gravitational waveforms over a time scale of about 20 ms after
  the merger for the V18 and SFHo EOSs obtained for gravitational masses
  $2\times1.35\ms$, comparing different choices of constant $\gt=1.1,
  1.5, 1.7, 1.75$ and the FT EOSs. }
\label{f:strain}
\end{figure*}

\section{Numerical results}
\label{s:res}

In the following we present the results of our binary NS
mergers simulations. Technical details regarding the extraction of the
gravitational-wave signal are given in the Appendix.

\subsection{Bulk dynamics}

Following the considerations made in Sec.~\ref{s:eos} for the V18 EOS and
the chosen total binary mass $2.7\ms$, the merger simulations do not
feature an immediate collapse to a black hole, but produce a metastable
HMNS up to the largest time $t\approx20\,{\rm ms}$ that we reached in the
simulations. At that time, the remnant is still stabilized by
differential rotation and finite temperature contributions to the
pressure. This feature seems to be compatible with the multimessenger
analysis of the GW170817 event \cite{Gill2019}. On the other hand, the
simulations performed with the SFHo EOS lead to a rather rapid collapse
into a black hole, which seems to be in contrast with the expected amount of
mass ejected in the GW170817 event.

Figure~\ref{f:maxrhot} shows in the two top panels the evolution of the
maximum rest-mass density, $\rho_\text{max}$,
for the different cases we have studied,
while the third and fourth panel report the evolution of both the maximum
and the density-weighted average temperature
\begin{equation}
 \expval{T} \equiv \frac{\int dV\rho\, T}{\int dV\rho} \:,
\label{e:tav}
\end{equation}
where the average is performed on the $z=0$ plane,
after applying a low-density threshold of $10^{13}\gc3$
to avoid contamination from the very light but very hot matter ejected;
only for the SFHo case,
we change this threshold to $10^{10}\gc3$ in order to also calculate
the averaged quantities even after the collapse.
A lighter color is chosen for the inspiral phase,
where such temperatures are meant as representative only and do not reflect an
accurate description of the thermodynamics of the matter.
A similar behavior (and even larger inspiral temperatures)
has been found also for other temperature-dependent EOSs, e.g.,
Refs.~\cite{Most2018b,Most2019b,Perego2019}.
In the lowest panel we also show for both FT EOSs
the density-weighted average relative deviation from beta stability
\begin{equation}
 \expval{ \frac{\Delta Y_{e\beta}}{Y_e} } \equiv
 \frac{\int dV \rho\ \frac{|Y_{e\beta}-Y_e|}{Y_{e\beta}} }{\int dV\rho} \:,
\label{e:dyav}
\end{equation}
where $Y_{e\beta}(\rho,T)$ represents the electron fraction
calculated pointwise on the $z=0$ plane assuming beta equilibrium
at the density $\rho$ and temperature $T$ of each point.
For the V18 EOS it stabilizes at a fairly large reduction of about 40\%,
which will be discussed later in more detail.
We set our time coordinate such that $t = t_{\rm merg} = 0$,
where $t_{\rm merg}$ is the time of the merger and
corresponds to the maximum of the gravitational-wave amplitude,
for all the cases we study.

When considering V18-EOS simulations, we find that, unsurprisingly, the
$\gt=1.1$ simulation produces the remnant with the highest maximum
rest-mass density ($\rho_{\rm max}\approx 10^{15}\gc3$), which decreases
to about $0.94\times10^{15}\gc3$ with increasing $\gt$. Indeed, this is
simply the consequence of the fact that increasing the thermal support
against gravity leads to a less dense remnant. Interestingly, the
temperature-dependent EOS leads to a remnant with an even smaller maximum
rest-mass density ($\rho_{\rm max}\approx 0.88\times10^{15}\gc3$) than
the hybrid-EOS cases.
This feature points to a systematic difference
between the two types of simulations:
while the hybrid method is by construction based on an EOS of
cold betastable matter with thermal corrections,
the full simulation produces matter strongly out of beta equilibrium,
see the lowest panel of Fig.~\ref{f:maxrhot},
as will be analyzed later.

On the other hand, the simulations carried out with the SFHo EOS show
that the remnant collapses into a black hole after a time which strongly
depends on the chosen thermodynamical treatment. In particular, the
collapse takes place at $t\approx13\,{\rm ms}$ for the \FT EOS and at
$t\approx7\,{\rm ms}$ or $t\approx14\,{\rm ms}$ for the cases in which
$\gt=1.50$ or $\gt=1.75$, respectively.
Furthermore, before collapse, the
fluctuations of the rest-mass density and temperature are more violent
than for the V18 EOS during this metastable phase.
While we focus here on the dependence of the collapse time
on the temperature treatment,
it has also been found to depend sizeably on the numerical resolution (see, e.g. Refs.
\cite{Kiuchi2014,DePietri2020}),
which we have not been able to study here due to lack of numerical resources.

\begin{figure*}[t]
\vspace{-0mm}\hspace{6mm}
\includegraphics[scale=0.28]{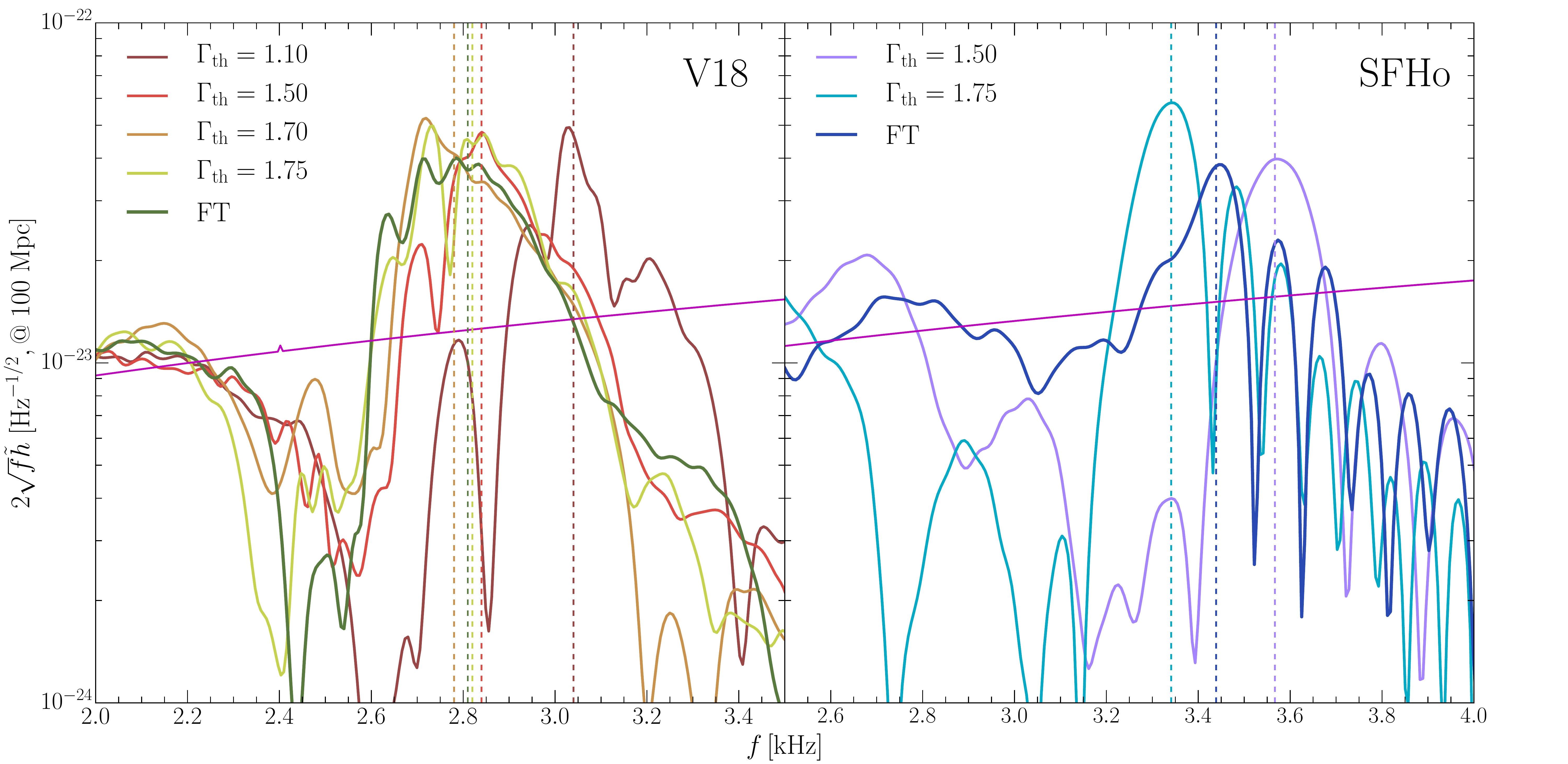}
\vspace{-3mm}
\caption{ PSDs of the simulations with the V18 and SFHo EOSs, at a
  distance of 100 Mpc. Vertical dashed lines of different colors
  indicate the frequency $f_2$. The sensitivity curve (magenta color) of
  Advanced LIGO is displayed for reference. }
\label{f:psd}
\end{figure*}

\begin{figure*}[t]
\vspace{-0mm}
\centerline{\hspace{4mm}
\includegraphics[scale=0.22]{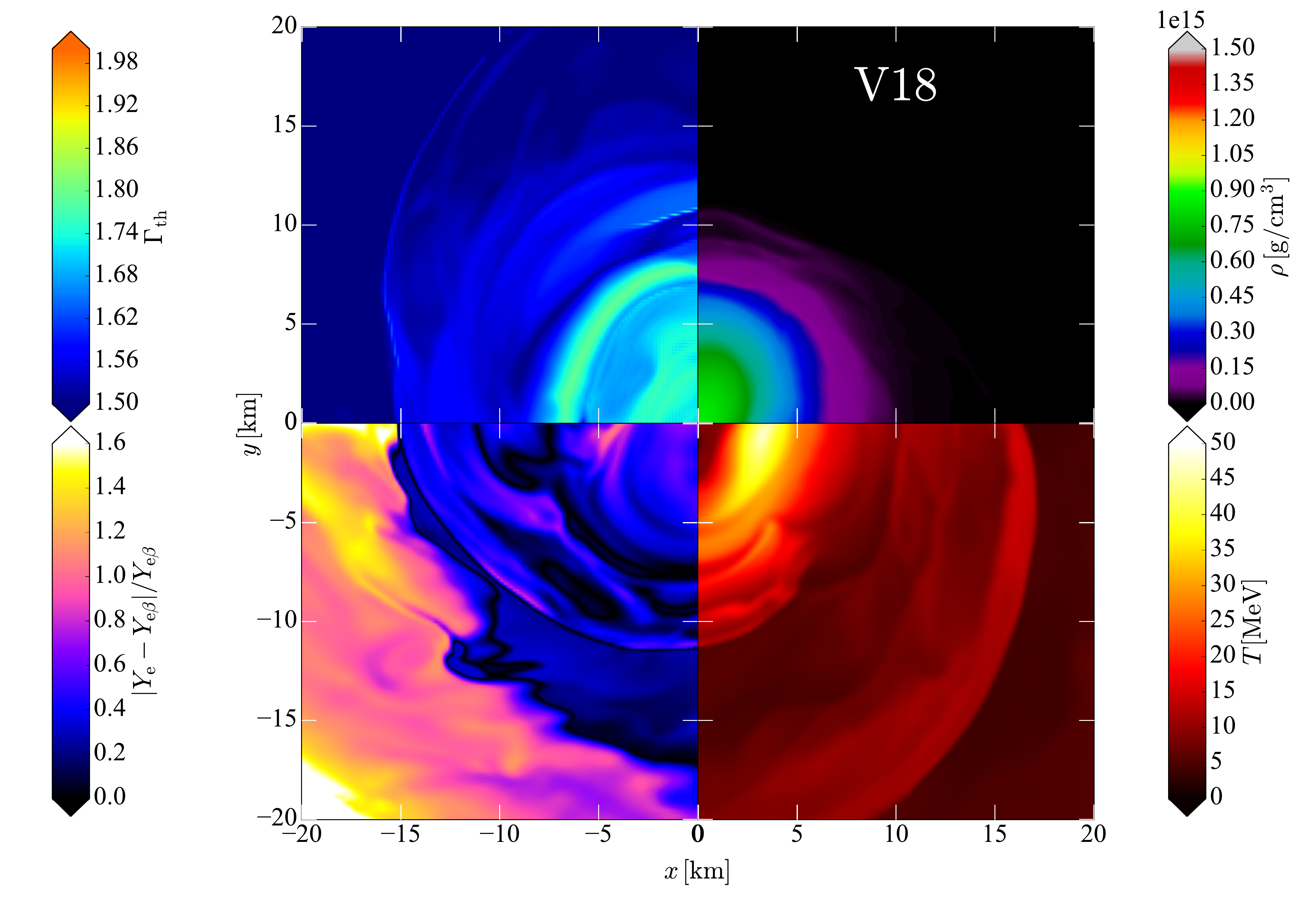}\includegraphics[scale=0.22]{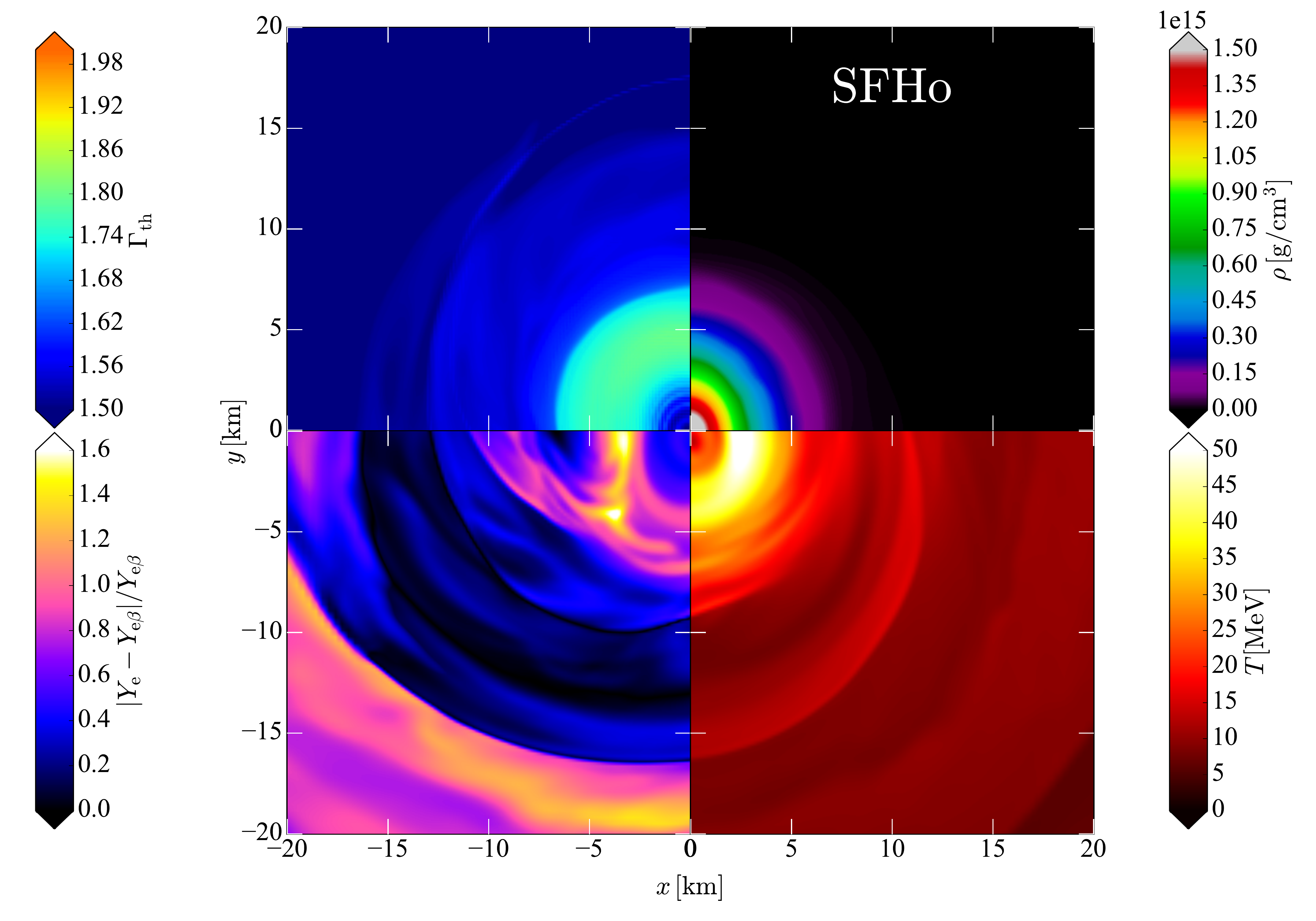}}
\vspace{-3mm}
\caption{ Distributions of $\gt$, Eq.~(\ref{e:gth}), (upper left side of
  the figures), rest-mass density (upper right), temperature (lower
  right), and deviation from beta stability (lower left), in the $z=0$
  plane at $t\approx9\,{\rm ms}$ after the merger. }
\label{f:combinedgamma}
\end{figure*}

As mentioned previously, in addition to the maximum temperature for the
\FT simulations, whose values during the postmerger phase peak at around
$70\mev$ and $110\mev$ for the V18 and SFHo EOS
respectively, we also report the density-weighted average temperature. Note
that for both EOSs, even during the inspiral, the average temperature is
much smaller than the maximum values, which, especially during the
inspiral, are reached only in small zones of the computational domain, as
will be illustrated in Fig.~\ref{f:combinedgamma}.

We also confirm that,
while during the inspiral phase there is no great deviation from beta-stability,
with average values mostly below 5\%,
the post-merger remnant manifests important differences with respect to the latter,
with average deviations of about 40\% and 50\% for V18 and SFHo respectively.

\subsection{Gravitational-wave emission}

In Fig.~\ref{f:strain} we show the plus polarization of the $\ell=m=2$
component of the gravitational-wave strains, which we label as
$h^{22}_+$, for all the considered simulations we have carried out using
the V18 and SFHo EOSs. As expected, no significant differences are found
in the inspiral part of the signal, the only notable feature being
that the time of merger, which we consider as the time corresponding to
the maximum of the strain amplitude, varies slightly when varying $\gt$
in the hybrid EOS approach (the maximum variations are about
$0.05\,{\rm ms}$ with respect to the average times calculated for both
EOSs in the hybrid EOS approach). The time of merger measured in the FT
runs for both the V18 and the SFHo EOS differs instead of $\approx
0.6\,{\rm ms}$ with respect to the average time calculated in the
hybrid-EOS approach; we believe the small difference arises from the fact that while in the hybrid EOS approach finite temperature effects during the inspiral are minimized, the FT approach leads, especially in the final parts of the inspiral, to a temperature increase which, together with the slight deviation of beta-stability, could be responsible for this feature. On the other hand, as clearly shown in
Fig.~\ref{f:strain}, we find that all the cases considered here exhibit
very different postmerger profiles.

Figure~\ref{f:psd} shows the power spectral density (PSD)
plots of all simulations,
determined as detailed in the Appendix.
In particular, we choose to study the dominant
$\ell=m=2$ mode, and consider the position of the $f_2$ peak (following
the same nomenclature as in Ref.~\cite{Rezzolla2016}) as a tracker of the
different behaviors. Since especially for the V18-EOS case with higher
$\gt$ it is difficult to distinguish the dominant $f_2$ peaks, a fitting
procedure represents the only way for an accurate determination of the
$f_2$ positions (see the Appendix for a discussion on the determination
of the values of the peaks).
We report in Table~\ref{t:gt} these values,
together with their indetermination, the $f_\text{max}$ values for each
simulation, and the emitted gravitational-wave energy $E_\text{GW}$ for
the $\ell=m=2$ mode, measured as outlined in the Appendix. In general,
$f_2$ decreases and $E_\text{GW}$ increases with increasing $\gt$, while
the values of $f_\text{max}$ depend only very weakly on $\gt$ and do not
show any specific dependence.
As a result, and accounting for the fact
that the determination of the $f_2$ peak frequency inevitably comes with
a considerable uncertainty related to the different distributions of
power in the various PSDs, the only robust conclusion that can be drawn
from the data in Table \ref{t:gt} is that values of the thermal adiabatic
index such that $\gt < 1.5 $ are not in agreement with the
results of the \FT simulations. In the following sections we will seek
other and more robust indicators of the optimal value for $\gt$.

\begin{table}[b]
\caption{
GW properties of NSs for the considered EOSs:
instantaneous frequency at amplitude maximum $f_{\rm{max}}$,
frequency of the $f_2$ peak,
and the total emitted energy $E_{\rm{GW}}$.}
\def\myc#1{\multicolumn{1}{c}{$#1$}}
\renewcommand{\arraystretch}{1.2}
\begin{ruledtabular}
\begin{tabular}{@{}ldld}
 Simulation
 & \myc{f_\text{max}\;[\text{kHz}]}
 & \myc{f_\text{2}\;[\text{kHz}]}
 & \myc{E_\text{GW}\;[\text{$10^{52}$ erg}]} \\
\hline 
 V18 - \FT               & 1.77 & 2.81$\pm$0.02 & 5.28\\
 V18 - $\gt=1.75$        & 1.79 & 2.82$\pm$0.08 & 5.84\\
 V18 - $\gt=1.7$         & 1.77 & 2.78$\pm$0.07 & 5.68\\
 V18 - $\gt=1.5$         & 1.79 & 2.84$\pm$0.01 & 4.97\\
 V18 - $\gt=1.1$         & 1.81 & 3.04$\pm$0.01 & 4.46\\
 SFHo - \FT              & 1.95 & 3.44$\pm$0.01 & 6.89\\
 SFHo - $\gt=1.75$       & 1.92 & 3.34$\pm$0.01 & 7.80\\
 SFHo - $\gt=1.5$        & 1.93 & 3.57$\pm$0.01 & 6.38\\
\end{tabular}
\end{ruledtabular}
\label{t:gt}
\end{table}

\begin{figure*}[t]
\centerline{\includegraphics[scale=0.28]{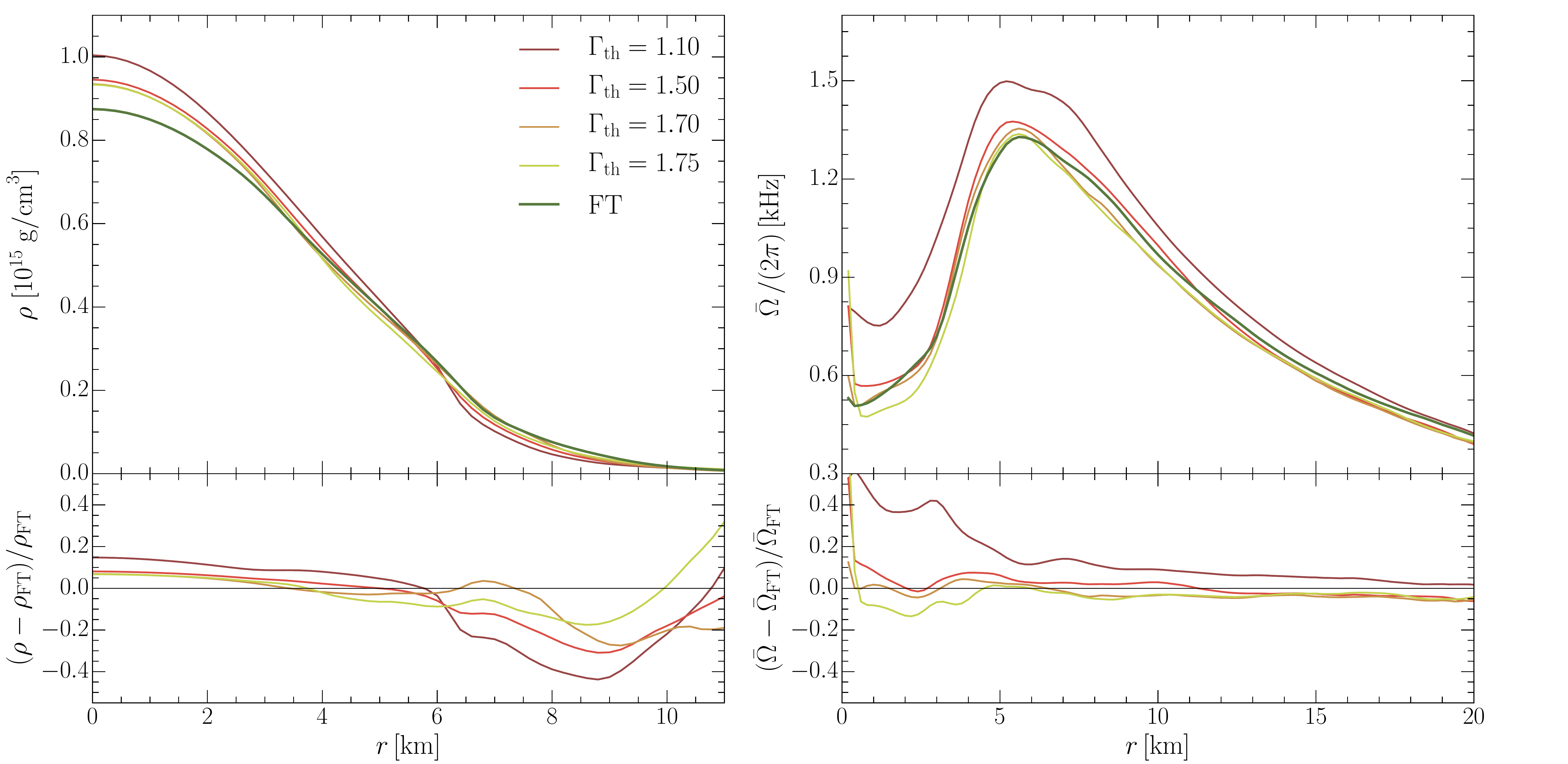}}
\vspace{-3mm}
\caption{ Average rest-mass density and angular velocity for the
  different V18-based EOSs as a function of the radial coordinate $r$
  ($z=0$) at $t=14\,{\rm ms}$. }
\label{f:rho_prof}
\end{figure*}

We further note that the values of the $f_2$ frequencies reported in
Table \ref{t:gt} agree reasonably well with both the universal relation
between $f_2$ and the tidal polarizability parameter $k_2^T$
\cite{Rezzolla2016} and the radius of a 1.6$\ms$ NS, $R_{1.6}$
\cite{Bauswein2012}, which we report for completeness:
\begin{align}
& f_2 \approx 5.832-1.118\, \left(k_2^T\right)^{1/5} \approx 2.95 \ &[\text{kHz}] \,,\\
& f_2 \approx 8.713 - 0.4667\, R_{1.6} \approx 2.86               \ &[\text{kHz}] \,,
\end{align}
where $k_2^T = 113.08$, $R_{1.6}=12.54\,{\rm km}$ for the V18 EOS, while
$k_2^T = 78.75$, $R_{1.6}=11.77\,{\rm km}$ for the SFHo EOS.

We also find that the simulation employing the V18 EOS with $\gt=1.1$
yields the highest frequency of the $f_2$ peak ($\sim230\,{\rm Hz}$
above the \FT value). Such a finding is in agreement with the behavior of
the rest-mass density found in Fig.~\ref{f:maxrhot}. In particular, since
the frequency of the mode scales with the square root of the average
density (see, e.g., Ref.~\cite{Kokkotas99b}), the behavior of the $f_2$
peak confirms spectroscopically that in this case the remnant not only
has the largest central density, but it also has the largest average
rest-mass density and is therefore subject to the fastest
oscillations among all the cases considered.

\begin{figure}[t]
\vspace{-6mm}
\centerline{\includegraphics[scale=0.37]{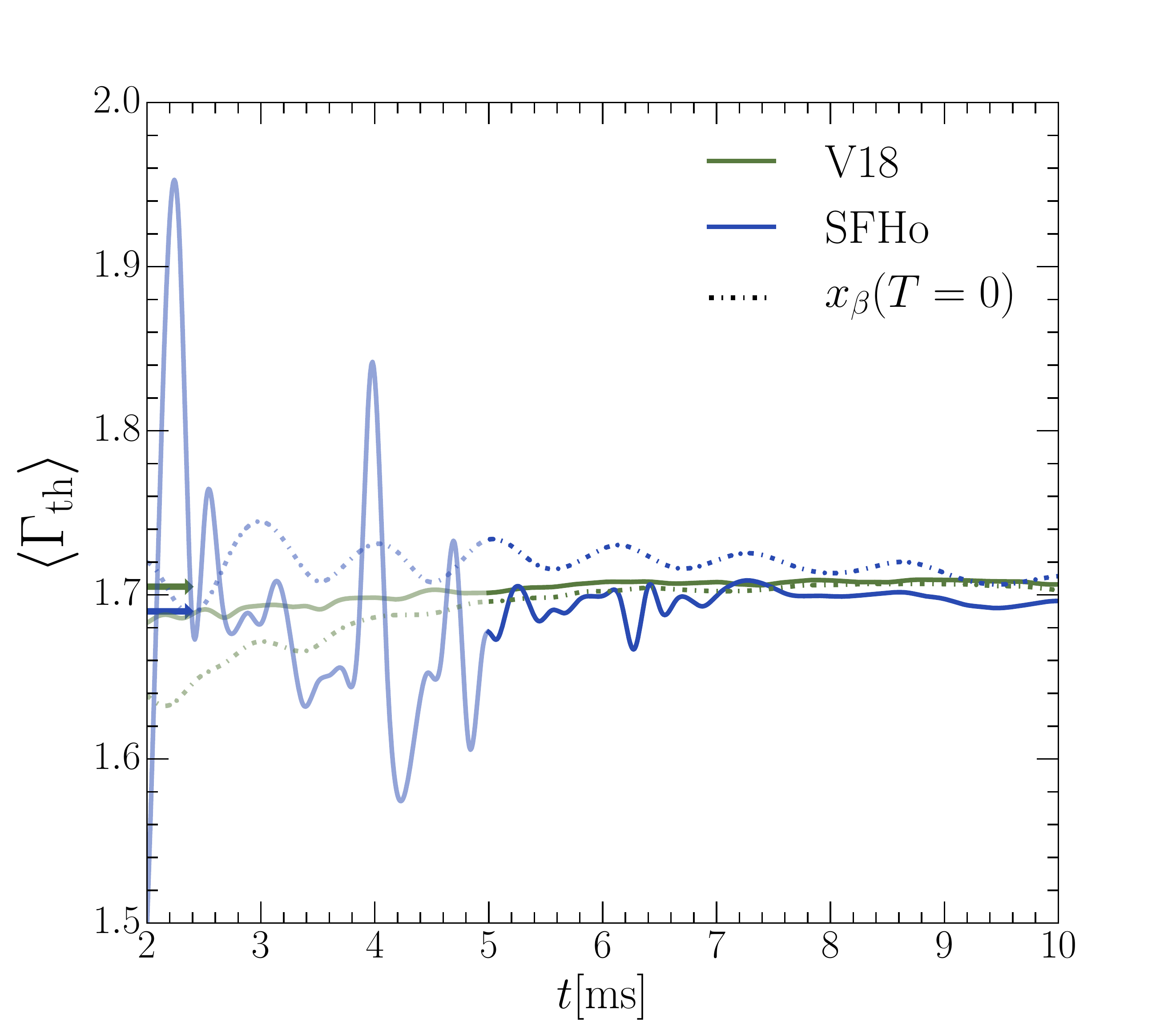}}
\vspace{-2mm}
\caption{
The average $\gt$, Eq.~(\ref{e:gav}),
as a function of time for the \FT V18 and SFHo simulations.
Time averages related to the total time interval considered here
are represented as arrows in the plot.
Dotdashed curves represent the average values of $\gt$ calculated using
Eq.~(\ref{e:gth}) with $x_p=x_\beta(\rho,T=0)$,
as described in the text.}
\label{f:gammaprof}
\end{figure}

\begin{figure*}[t]
\vspace{-3mm}
\centerline{\includegraphics[scale=0.25]{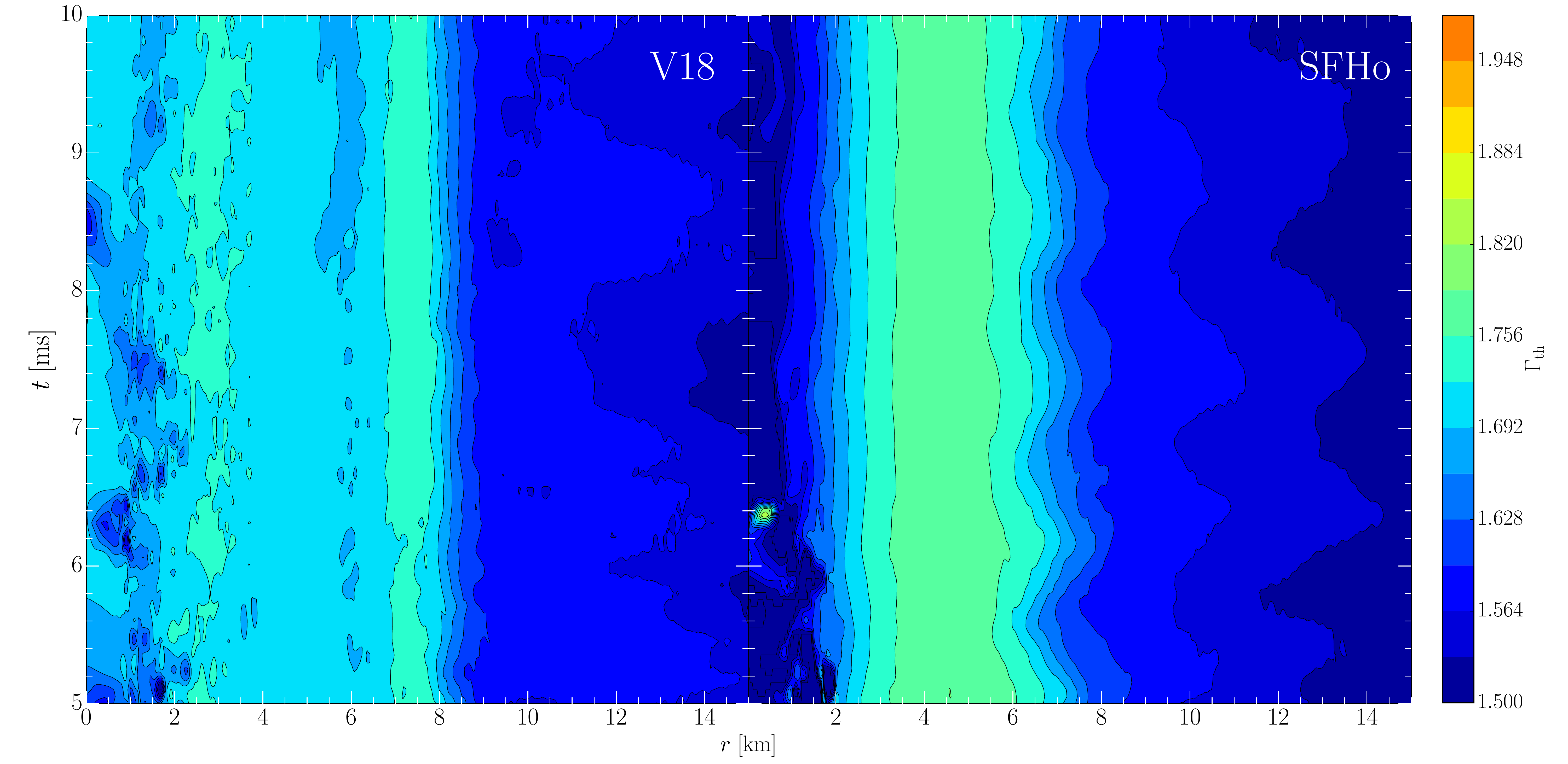}}
\vspace{-2mm}
\caption{
Iso-contours of $\gt$ as a function of cylindrical radius $r$ and
time $t$ for the V18 and SFHo \FT simulations.}
\label{f:gammacontour}
\end{figure*}

\subsection{Differential rotation and effective thermal adiabatic index}

In the following we analyze in more detail properties of the remnant that
is formed after merger. Figure~\ref{f:rho_prof}, in particular, shows the
one-dimensional profiles of the averaged rest-mass density (left panel)
and of the angular velocity (right panel) for all the cases we have
considered at a time $t\approx 14\,{\rm ms}$ after the merger. The
profiles are obtained from the values of the corresponding quantity on
the equatorial plane ($z=0$) and after averaging in the azimuthal direction
and over a time window of $1\,{\rm ms}$ so as to obtain functions that
depend only on the cylindrical radius, $r$, from the center of the grid.

In the bottom part of each panel we also report the fractional
differences of the hybrid-EOS profiles with respect to the fiducial \FT
ones. Overall, we find that in the core of the remnant (i.e., $r\lesssim
6\,{\rm km}$), differences in density remain below $10\%$ for the cases
$\gt=1.5,1.7,1.75$, while they increase below
$\rho\approx2\times10^{14}\gc3$. Interestingly, the case $\gt=1.1$ always
shows the largest differences and the case $\gt=1.7$ the smallest
fractional differences in the core area, which is dynamically the most
important one.

In order to determine which values of $\gt$ best approximate the \FT behavior,
we compute such values pointwise according to Eq.~(\ref{e:gth}),
using the local values of $\rho$, $\xp$, and $T$ obtained in the \FT simulations
and the \FT tables to compute $p$ and $\eps$.
Note, however, that while $\gt$ is used in simulations where the
betastability is enforced throughout the evolution,
this way of computing $\gt$ ignores the betastability condition of cold matter,
since $\xp$ --~which is evaluated pointwise in the \FT simulations~--
is not the proton fraction of cold betastable matter.
The method is most close to the fixed-$\xp$ prescription used in Fig.~\ref{f:xt}
with $x_{\beta}(T>0)$,
but the \FT $\xp$ is not the one of hot betastable matter either.
As a result, it can only give an approximate indication
of the ``best'' value to be used in hybrid-EOS calculations
(see also the previous discussion in Sec.~\ref{s:gt}).

Figure~\ref{f:combinedgamma} shows in the top left quadrants the values
of the ``local'' $\gt$ on the $z=0$ plane at time $t=9\,{\rm ms}$ after
the merger. Other quantities reported are: the distributions of the
rest-mass density $\rho$ (top right quadrants), the temperature $T$
(bottom right quadrants),
and the deviation of the electron fraction from its betastable value,
\hbox{$(Y_{e\beta}-Y_e)/Y_{e\beta}$} (bottom left quadrants).
Note that while in the hybrid-EOS simulations $\gt$ is, by construction,
constant over the computational domain,
in the \FT case the computed value with the V18 EOS (right panel)
is generally $\gt \lesssim 1.6$ for $\rho \lesssim 5\times10^{13}\gc3$,
and very close to $\gt \simeq 1.7$ for higher
densities and hence in the core of the HMNS.

On the other hand, the SFHo-simulation (right panel) exhibits a slightly
different behavior, with $\gt \gtrsim 1.7$ in the density region $10^{14}
\lesssim \rho/\gc3 \lesssim 10^{15}$ and with the highest-density region
being instead dominated by values $\gt \lesssim 1.65$. This behavior
confirms qualitatively the conclusions drawn from Fig.~\ref{f:psd},
namely, that a value of the thermal adiabatic index $\gt \approx 1.7$
provides a good match to the post-merger spectroscopic properties
observed in the two FT EOSs.

The temperature distributions reported in Fig.~\ref{f:combinedgamma} show
the typical appearance of two hot spots of more than $50\mev$
\cite{Kastaun2014,Hanauske2016},
whose temperature evolution was shown in Fig.~\ref{f:maxrhot}
and whose appearance can be associated with the
conservation of the Bernoulli constant
(see \cite{Hanauske2016} for a detailed discussion).
The two hot spots eventually merge into an
axisymmetric structure after $t \simeq 22\,{\rm ms}$.
Also quite evident from the bottom-left quadrants is that
the matter after the merger is significantly out of beta equilibrium,
especially in the low-density layers of the HMNS.
Averaged values were plotted in Fig.~\ref{f:maxrhot}.
As discussed above, this deviation limits the
validity of the comparison of the dynamical and thermodynamical
properties of the matter between simulations carried out with the
FT EOSs and with hybrid EOSs.

It is also clearly visible from the top-left quadrants in
Fig.~\ref{f:combinedgamma} that the local value $\gt$ is far from being
constant, but depends strongly on density, temperature, and proton
(electron) fraction at each point of the computational domain.
Notwithstanding these limitations, we can nevertheless attempt to
identify in \FT simulations a reference value of $\gt$ by considering a
spatial average and by inspecting how much this average varies with time.
For this purpose we calculate, on the equatorial plane ($z = 0$) and at each time $t$ after the merger,
the density-weighted spatial average of $\gt$ as
[cf.~Eq.~(\ref{e:tav}) for the densty-weighted average temperature]
\begin{equation}
\expval{\gt} \equiv \frac{\int d{V}\rho\gt}{\int d{V}\rho} \:,
\label{e:gav}
\end{equation}
where, again, the average is performed after applying a low-density
threshold of $10^{13}\gc3$ to avoid contamination from the dynamically
unimportant matter. We have verified that the results are insensitive to
changes of this limit threshold, with deviations of $\gt$ of the order
$3\times10^{-3}$ when $10^{12}\gc3$ is chosen instead.

Figure~\ref{f:gammaprof} reports the evolution of the average thermal
adiabatic index,
in a time window between $5$ and $10\,{\rm ms}$ after merger,
which corresponds to the time interval when the fluctuations of
$\gt$ for the SFHo EOS are minimal and a comparison between FT and hybrid
EOSs is more reasonable.
We notice that, for both EOSs, $\expval{\gt}\simeq 1.7$,
and that the corresponding time and spatial averages
for the V18 and the SFHo EOS are
$\expval{\bar{\Gamma}_{\rm th}}=1.705$ and
$\expval{\bar{\Gamma}_{\rm th}}=1.690$, respectively
(indicated by arrows).
These averages include also the initial time interval,
$2 \lesssim t/{\rm ms} \lesssim 5$,
when the HMNS has just been formed
and the dynamics is still very far from being quasi-stationary
(light-colored curve segments).
As a further confirmation of our results,
we also report the average of $\gt$ calculated using
the values of $p$ and $e$ evaluated employing $x_{\beta}$ at $T=0$
(instead of the local value of $x_p$),
as we have done in Fig.~\ref{f:gam} (dashdotted curves).
We find also in this case good agreement with the value $1.7$ for both EOSs.

Figure~\ref{f:gammacontour} shows a selection of $\gt$ iso-contours
on the $z=0$ plane for the time window between $5$ and $10\,{\rm ms}$
also considered for Fig.~\ref{f:gammaprof}. We find that the
distribution shown in Fig.~\ref{f:combinedgamma} remains robust for the
time window considered; in particular, for both V18 and SFHo the $\gt$
distribution peaks off-centre. We notice that V18 is characterized
by two stable and narrow peak-structures at about $3$ and $7$ km, while
SFHo shows a broader peak-region, approximately comprised between
$3$ and $6$ km. The high density regions also show important differences,
being characterized by higher $\gt$ for V18 and values even lower than $1.5$
for SFHo. The latter case shows local strong oscillations about the center
which are evident for the first ms of the time-window we show, representing
a residual of the stronger oscillations affecting the previous part of the
simulation.

In summary, on the basis of the various measurements and diagnostics
discussed so far, we conclude that using a hybrid EOS to
simulate the merger of binary NS systems, the value of thermal
adiabatic index $\gt\approx 1.7$ best approximates the dynamical and
thermodynamical behavior of matter computed using complete,
finite-temperature EOSs.

\section{Conclusions}
\label{s:end}

Hybrid EOSs, in which thermal contributions are artificially added in
terms of an ideal-fluid EOS, are widely adopted in the numerical
modelling of merging binary NSs. This is in part due to the
smaller computational costs that are associated with hybrid EOSs, but,
more importantly, it is the consequence of the scarcity of full
temperature-dependent EOSs that can be employed in numerical
simulations. The use of such hybrid EOSs, however, also raises the
fundamental problem of deciding which value should be given and kept
constant -- both in space and time -- to the thermal adiabatic index
$\gt$, which is instead expected to change both in space and time.

In order to address this point, and hence determine the optimal value for
$\gt$, we have carried out a number of simulations of merging neutron
stars in full general relativity, employing two fully tabulated,
temperature-dependent EOSs and a neutrino-leakage scheme for the treatment
of neutrinos. The first of these temperature-dependent EOSs, the V18 EOS,
has been derived in the BHF formalism that fulfils all the current
constraints imposed by the nuclear phenomenology, and also respects
recent observational limits on the maximum NS mass and
deformability; the V18 EOS has been employed here for the first time in
merger simulations. The second temperature-dependent EOS, the SFHo EOS,
is based on a RMF model which takes into account a statistical ensemble
of nuclei and interacting nucleons; the SFHo EOS has been employed
routinely in the past to model merging NS binaries.

Together with the temperature-dependent EOSs, we have also performed
similar simulations employing hybrid EOSs where we have considered a
variety of values for the thermal adiabatic index $\gt$ and
where the cold part is given by the slice at $T=0$ of the
temperature-dependent EOSs. In this way, we were able to construct
instances of the binaries that were virtually the same during the
inspiral -- when thermal effects are dynamically unimportant -- and that
start to differ from the merger, as the thermal contributions from the
two classes of EOSs are important and different.

We have then used and monitored a number of different quantities relative
either to the matter sector -- e.g., rest-mass density, temperature,
electron fraction, angular velocity of the merged object -- or to the
gravitational-field sector -- e.g., gravitational waves and PSDs of the
post-merger signal. Furthermore, we have performed measurements of the
effective thermal adiabatic index and followed its distribution in space
and its evolution in time.
The importance of ambiguities in its definition due to the loss of
beta equilibrium during the postmerger simulation have been evidenced.
In this way, and collecting the information from all of these quantities,
we conclude that a value of $\gt\approx 1.7$ best approximates the complete,
finite-temperature EOS in binary NS simulations.
This value is similar to the standard one employed in numerical simulations
so far (i.e., $\gt = 1.75-1.80$),
but also importantly lower.
Future work will be aimed at increasing the robustness of this finding
by employing other temperature-dependent EOSs,
including those presented recently in Ref.~\cite{Lu2019}.

\section*{Acknowledgements}

We wish to express particular thanks to L.~Rezzolla
for the valuable support and discussions that made this work possible.
We also acknowledge useful discussions with K.~Takami and R.~De Pietri and thank D.~Radice for
the help and support with \texttt{WhiskyTHC}. Partial support comes from
``PHAROS,'' COST Action CA16214. Simulations have been carried out on the
\hbox{MARCONI} cluster at CINECA, Italy, on the Goethe cluster at CSC in
Frankfurt, and on the SuperMUC cluster in Munich.
Support from INFN ``Iniziativa Specifica NEUMATT'' is also acknowledged.
Part of this work made use of the computational resources provided under
project ``Digitizing the universe: precision modelling for precision cosmology'',
funded by the Italian Ministry of Education, University and Research (MIUR).
This work is also
sponsored by the National Natural Science Foundation of China under Grant
Nos.~11475045, 11975077 and the China Scholarship Council,
No.~201806100066.

\section*{Appendix}
\label{s:app}

\subsection{Gravitational-wave signal}
\label{sec:gws}

We extract the gravitational-wave signal using the standard
Newman-Penrose formalism \cite{Newman62a}: we calculate the
Newman-Penrose scalar $\psi_4$ at different surfaces of constant
coordinate radius $r$ using the Einstein Toolkit module
\textsc{WeylScal4}. In particular, $\psi_4$ is related to the second time
derivatives of the gravitational-wave polarization amplitudes $h_+$ and
$h_\times$ by
\begin{equation}
 \psi_4 = \ddot{h}_+ - i\ddot{h}_\times =
  \sum_{l=2}^\infty \sum_{m=-l}^l \psi_4^{\ell m}(t,r)\,_{-2}Y_{\ell m}(\theta,\phi)
\,,
\end{equation}
where the double dot represents the second time derivative and we have
introduced also the multipole expansion of $\psi_4$ in spin-weighted
spherical harmonics \cite{Goldberg:1967} of spin weight $s=-2$ (such
decomposition is performed by the module \textsc{Multipole}). As the
dominant mode is $\ell=m=2$, we restrict our analysis only to the latter,
i.e., we assume
\begin{equation}
 h_{+,\times} = \sum_{l=2}^\infty \sum_{m=-l}^l h_{+,\times}^{\ell
   m}(t,r)\,_{-2}Y_{\ell m}(\theta,\phi) \approx
 h^{22}_{+,\times}\,_{-2}Y_{22}(\theta,\phi) \,.
\label{eq:hpol}
\end{equation}
The fixed-frequency integration described in \cite{Reisswig:2011} is
carried out in order to double integrate $\psi_4$ in time. We then align
our waveforms, as in \cite{Rezzolla2016}, to the ``time of the merger,"
which we set as $t=0$ and we define as the time when the GW amplitude
\begin{equation}
 |h| \equiv \sqrt{h^2_+ + h^2_\times}
\label{eq:hmod}
\end{equation}
is maximal. We also compute the phase of the complex waveform, which we
label with $\chi$ = arctan$(h_\times / h_+)$, and the instantaneous
frequency of the gravitational waves, defined as in \cite{Read2013},
\begin{equation}
 f_\text{GW} \equiv \frac{1}{2\pi}\dv{\chi}{t} \,.
\end{equation}
We identify, as in \cite{Rezzolla2016},
$f_\text{max} \equiv f_\text{GW}(t=0)$
as the instantaneous frequency at amplitude maximum.

The total emitted energy for the $\ell=m=2$ mode is
\begin{equation}
 E_\text{GW} = \frac{R^2}{16\pi} \int {dt}\int {d\Omega}\;
 \abs{\dot{h}(t,\theta,\phi)}^2 \,,
\end{equation}
where $\Omega$ is the solid angle
and $R$ represents the source-detector distance.

We also consider the power spectral density (PSD) of the effective
amplitude, defined as
\begin{equation}
 \tilde{h}(f) \equiv
 \sqrt{\frac{\abs{\tilde{h}_+(f)}^2 + \abs{\tilde{h}_\times(f)}^2}{2}} \,,
\end{equation}
where $\tilde{h}_{+,\times}(f)$ are the Fourier transforms of $h_{+,\times}$,
\begin{equation}
 \tilde{h}_{+,\times}(f) \equiv
 \int {dt}e^{-i2\pi ft} h_{+,\times}(t)
\end{equation}
for $f \geq 0$, and $\tilde{h}_{+,\times}(f) \equiv 0$ for $f<0$. We
determine the position of the $f_2$ peak of the PSD, after applying a
symmetric time-domain Tukey filter with parameter $\alpha=0.25$ to the
waveforms, in order to compute PSDs without the artificial noise due to
the truncation of the waveform. We then fit our data with the analytic
function \cite{Takami2015}
\begin{equation}
 S_2(f) = A_{2G} e^{-(f-F_{2G})^2/W_{2G}^2} + A(f) \gamma(f) \,,
\end{equation}
where
\begin{align}
 A(f) &\equiv \frac{1}{2W_2}
 \qty[ (A_{2b}-A_{2a})(f-F_2) + W_2(A_{2b}+A_{2a}) ] \,,
\\
 \gamma(f) &\equiv \qty( 1+e^{-(f-F_2+W_2)/s} )^{-1}
 \qty( 1+e^{(f-F_2-W_2)/s} )^{-1} \,.
\end{align}
The peak frequency is then determined by
\begin{equation}
 f_2 \equiv \frac{\int {df}\,S_2(f)\,f}{\int {df}\,S_2(f)} \,.
\end{equation}

This fitting procedure manifests an intrinsic uncertainty due to both the
choice of the fitting functions and parameters, and the integration
interval, which we estimate as $\pm 10\,\rm{Hz}$. Such indetermination is
later added in quadrature to a systematic deviation of the value we find
for $f_2$ from the nearest (local) maximum of the PSD curve. The latter
estimate coincides with the deviation with respect to the global maximum
of the PSD for all the cases considered apart from the $\gt = 1.75$ case,
where the presence of a second narrower peak located at lower frequencies
determines a higher indetermination. The case $\gt = 1.7$ also shows the
same feature, with the two peaks being indistinguishable with respect to
each other. Table~\ref{t:gt} reports the total indetermination for each
case, namely, the sum in quadrature of the intrinsic uncertainty and the
deviation with respect to the global maximum of the PSD curves.
\newcommand{\nphysa}{Nuclear Physics A}
\bibliographystyle{apsrev4-1-noeprint}
\bibliography{aeireferences} 

\end{document}